\documentclass[useAMS,usenatbib,usegraphicx]{mn2e}

\newcommand{\der}{\mathrm{d}}

\title[Mapping stationary axisymmetric phase-space distribution functions by orbit libraries]{Mapping stationary axisymmetric phase-space distribution functions by orbit libraries}
\author[J. Thomas, R. P. Saglia, R. Bender, D. Thomas, K. Gebhardt, J. Magorrian, D. Richstone]{J. Thomas$^{1}$\thanks{E-mail:jthomas@mpe.mpg.de}, R. P. Saglia$^{1,2}$, R. Bender$^{1,2}$, 
D. Thomas$^{2}$, K. Gebhardt$^{3}$, 
\newauthor J. Magorrian$^{4}$, D. Richstone$^{5}$\\
$^{1}$Universit\"atssternwarte M\"unchen, Scheinerstra\ss e 1, D-81679 M\"unchen, Germany\\
$^{2}$Max-Planck-Institut f\"ur Extraterrestrische Physik, Giessenbachstra\ss e, D-85748 Garching, Germany\\
$^{3}$Department of Astronomy, University of Texas at Austin, C1400, Austin, TX78712, USA\\
$^{4}$Theoretical Physics, Department of Physics, University of Oxford, 1 Keble Road, Oxford U.K., OX1 3NP\\
$^{5}$Department of Astronomy, Dennison Bldg., University of Michigan, Ann Arbor 48109}
\begin{document}

\date{Accepted 1988 December 15. Received 1988 December 14; in original form 1988 October 11}

\pagerange{\pageref{firstpage}--\pageref{lastpage}} \pubyear{2002}

\maketitle

\label{firstpage}

\begin{abstract}
This is the first of a series of papers dedicated to unveil the mass composition and dynamical
structure of a sample of flattened early type galaxies in the Coma cluster. We
describe our modifications to the Schwarzschild code of Richstone et al. (in preparation). 
Applying a Voronoi tessellation in the surface of section we are able to assign
accurate phase-space volumes to individual orbits and to reconstruct the
full three-integral phase-space distribution function (DF) of any axisymmetric orbit library. 
Two types of tests have been performed to check the accuracy with which DFs can be
represented by appropriate orbit libraries. First, by mapping DFs of spherical 
$\gamma$-models and flattened Plummer models on to the library we show that the 
resulting line-of-sight velocity distributions and internal 
velocity moments of the library match those directly derived from the DF to a precision 
better than that of present day observational errors. Second, by fitting libraries to 
the projected kinematics of the same DFs we show that the distribution function 
reconstructed from the fitted library matches the input DF to a rms of about 15 per cent over 
a region in phase-space covering 90 per cent of the mass of the library. The achieved accuracy
allows us to implement effective entropy-based regularisation to fit real, noisy and
spatially incomplete data.
\end{abstract}



\section{Introduction}

Since the pioneering work of \citet{S79} orbit superposition techniques have 
become an important tool in the dynamical modelling of spheroidal stellar systems. 
Stationary distribution functions (DFs) of such systems are subject to Jeans' theorem 
and hence depend on the phase-space
coordinates only via the integrals of motion. In the axisymmetric case these integrals
are energy $E$, angular momentum along the axis of symmetry $L_z$ and for most potentials
an additional, non-classical ``third integral'' $I_3$. Because any set of integrals of
motion essentially represents an orbit and vice versa, the DF can be approximated by the
sum of single-orbit DFs, with the only adjustable parameter being the total amount of light
carried by the orbit. The main task that remains to adequately describe hot stellar
systems is to find an appropriate set of orbits.

Orbit superposition techniques have been used to model such systems in various symmetries 
(e.g. \citealt{R97}; \citealt{roman97}; \citealt{vdM98}; \citealt{cretton99};
\citealt{cappellari02}; \citealt{verolme02b}; \citealt{G03}; \citealt{vandeven03}), with 
the goal of determining different dynamical parameters like central black hole masses, 
internal velocity anisotropy or global mass-to-light ratios. An orbit library tracing the 
phase-space structure of a trial potential is fitted to observed photometry and kinematics,
to decide, whether or not it gives a valid model of the corresponding galaxy.

In the spherical case there exists a well-known mass-anisotropy degeneracy permitting
in general convincing fits to the projected velocity dispersion $\sigma$ 
even if the trial potential 
differs from the true one \citep{BM82}. With complete knowledge of the 
full line-of-sight velocity distributions (LOSVDs) however, it is possible to reconstruct
the DF, given the potential is known \citep{DM92}. Furthermore, even for the realistic case
where the potential is not known in advance \citet{MS93} and \citet{Ger93} have shown
how the information contained in the LOSVDs can constrain both, the 
potential and the DF.

Likewise, in the axisymmetric case, \citet{DG93} have calculated realistic smooth DFs and 
have shown that a similarily close relationship exists between the potential and internal 
kinematics on the one side and the projected kinematics on the other side. However, 
fits of axisymmetric libraries still pose some additional open questions. 
Recently, \citet{valluri04} discussed the indeterminacy of the reconstruction of the
potential in general axisymmetric systems from two- or three-dimensional data sets,
by studying the shape of the $\chi^2$-contours describing the quality of the orbital fit.
\citet{cretton04} argued
that even in case of a mathematically non-degenerate $f(E,L_z)$-system an artificial
degeneracy occurs, caused by the discreteness of the orbit library and emphasized the role
of appropriate smoothing, although not providing a definite solution. \citet{richstone04a}
critically analysed their arguments and emphasized that both high quality comprehensive
data sets and orbit libraries are needed to achieve a reliable modeling of axisymmetric
systems.

In view of this discussion concerning orbit-based dynamical models 
it seems worthwhile to step back and investigate how
well orbit libraries represent the phase-space structure of a given dynamical
system. This includes the examination of the choice of orbits, which in the generic 
axisymmetric case is difficult, since part of the phase-space structure is unknown due to
our ignorance related to $I_3$. A key tool for such an analysis are the
orbital phase-volumes which accomplish the transformation from the relative contributions 
of individual orbits to the library, the so called orbital occupation numbers or orbital 
weights, into phase-space densities (and vice versa). The availability of such phase-volumes 
actually offers several applications:
\begin{enumerate}
\item
\label{project df}
Accurate phase-volumes allow to 
calculate internal and projected properties like density and velocity profiles, 
line-of-sight velocity distributions (LOSVDs) etc. of general axisymmetric DFs $f(E,L_z,I_3)$
via orbit libraries. Besides the possibility to systematically study the structure of
general axisymmetric systems, these profiles provide a direct check of the choice of 
orbits via their comparison with those calculated from directly integrating the DFs.
\item
\label{reconstruct df}
From any fitted library one can reconstruct the corresponding DF via the phase-volumes and 
thus reconstruct the DF from any observed early-type galaxy in the axisymmetric approximation.
\item
\label{compare df}
If the library is fitted to some reference data constructed from a DF, then 
\ref{reconstruct df} 
allows to investigate how closely the fit matches the input DF and to implement an effective 
regularisation scheme, permitting to fit real (noisy) data sets.
\end{enumerate}

\citet{V84} touched the problem by writing down the transformation from cells of integrals
to the corresponding phase-space volumes. However, the resulting relations are only
suitable for explicitly known integrals, e.g. for single orbits only in the rare case all 
integrals are known. For components integrated about the unknown integrals they
have been exploited by e.g. \citet{R97}, \citet{cretton99} and \citet{verolme02a}.

The aim of the present paper is to introduce a general implementation for the calculation 
of individual orbital phase-volumes in any axisymmetric potential and, by following
applications \ref{project df} and \ref{compare df} to prove that our libraries accurately
map given dynamical systems. This directly justifies our setup of the library and 
sets the basis for our project to recover the dynamical structure and mass 
composition of a sample of flattened early-type galaxies in the Coma cluster. 
In a subsequent paper we will focus on the question of how much smoothing has to be applied 
in order to get an optimal estimate of the dynamical system underlying a given set of noisy 
and spatially incomplete observational data. The full analysis of the data set 
(\citealt{coma1}; \citealt{coma2}) will be addressed in a future publication.

The paper is organized as follows. In Sec.~\ref{library} we define all quantities related
to the library used in the subsequent Sections and describe our orbit sampling. 
Sec.~\ref{weights and densities} outlines the relation between orbital weights and orbital
phase-space densities. Sec.~\ref{phasevols} contains a description of our implementation
to calculate individual orbital phase-space volumes. In a first application we calculate
internal and projected properties of given DFs using orbit libraries in Sec.~\ref{projdf}.
Sec.~\ref{libfit} discusses how the library is fitted to given data sets and in
Sec.~\ref{dfrecon} we reconstruct reference DFs from their projected kinematics.
Finally, Sec.~\ref{conclusion} summarizes the results.


\section[]{The orbit library}
\label{library}
Our method to set up the orbit libraries used for the dynamical modelling
is based on the procedure presented in Richstone et al. (in preparation). There, the reader
finds a description of the basic properties of the program. In this
section we define quantities that are used later on in this paper.

In the following we assume that the luminosity density $\nu$ is known. 
In the analysis of real data it has to be obtained 
by deprojection of the measured photometry. With the stellar mass-to-light ratio
$\Gamma = (\frac{M}{L})$ the mass density $\rho_l$ of the luminous material follows
from $\nu$ as $\rho_l = \Gamma \, \nu$.

The total mass density $\rho$ possibly includes a dark component $\rho_\mathrm{DM}$ 
and reads
\begin{equation}
\rho = \rho \left( \Gamma,r_c,v_c \right) = 
\Gamma \, \nu + \rho_\mathrm{DM}
\end{equation}

Once the mass-profile is fixed, the potential $\Phi$ follows by integrating
Poisson's equation. With $\Phi$ known, a large set of orbits is calculated,
sampling homogeneously the phase-space connected with $\Phi$.

\subsection{Spatial and velocity binning}

\label{spatial grid}
As described in Richstone et al. (in preparation) we divide the meridional plane into
bins, equally spaced in $\sin \vartheta$\footnote{Throughout the paper, we use 
spherical coordinates ($r,\vartheta,\varphi$), with $\vartheta = 0\degr$ corresponding 
to the equatorial plane. If not stated otherwise, we use super- or subscripts $h,i,j,k$ 
as indices, $l,m,n$ as exponents.}, linear in $r$
near the inner boundary $r_\mathrm{min}$ of the library and logarithmic 
at the outer boundary $r_\mathrm{max}$,
(if not stated otherwise we use $N_r = 20$ radial bins, $N_\vartheta = 5$
angular bins.)
For the projection of the library we use
the same binning as for the meridional plane. Every spatial bin in the
plane of the sky is subdivided into $N_\mathrm{vel}$ bins linearily spaced
in projected velocities between $-v_\mathrm{max}$ and $v_\mathrm{max}$, leading to
a binsize for the LOSVDs of
\begin{equation}
\label{delta LOSVD}
\Delta v_\mathrm{LOSVD} = 2 \frac{v_\mathrm{max}}{N_\mathrm{vel}}.
\end{equation}
Even if the potential is spherical, then our spatial binning tags an axis
of symmetry. Later, when referring to a ``minor-axis'' we always mean
the axis $\vartheta = 90\degr$ of the library.

\subsection{Orbital properties}

\subparagraph*{Luminosity.}
For every orbit $i$ in the library we store its normalized contribution to the luminosity
$\der\mathrm{L}^j_i$ 
in spatial bin $1 \le j \le N_r \times N_\vartheta$, which equals the fraction of
time the orbit spends in bin $j$. Let $\Delta t^k_i$ denote the $k$-th timestep
in the integration of orbit $i$, so that
\begin{equation}
t^k_i \equiv \sum_{h \le k} \Delta t^h_i
\end{equation}
is the total time elapsed until timestep $k$ and 
\begin{equation}
{\cal J}^j \equiv \{ 
k : (r(t^k_i),\vartheta(t^k_i)) \in \mathrm{bin} \, j \}
\end{equation}
contains all timesteps during which orbit $i$ is located in spatial
bin $j$. Accordingly, we can write
\begin{equation}
\der\mathrm{L}^j_i = \sum_{k \in {\cal J}^j} \frac{\Delta t^k_i}{T_i},
\end{equation}
with $T_i \equiv \sum \Delta t^k_i$ being the total integration time of 
\mbox{orbit $i$}.

Given the orbit's weight $w_i$ to the whole library -- the
integrated luminosity along the orbit -- the total luminosity of the library 
in spatial bin $j$ reads
\begin{equation}
\label{def dL}
\der\mathrm{L}^j = \sum_i w_i \, \der\mathrm{L}^j_i.
\end{equation}

\subparagraph*{Internal velocity moments.}
To obtain the internal velocity moments
$\left< v^l_r v^m_\vartheta v^n_\varphi \right>$
of the orbit library, we store
for each orbit $i$ and each time step $\Delta t^k_i$ the product of velocities
$v^l_r v^m_\vartheta v^n_\varphi$ and fractional time
$\Delta t^k_i/T_i$. 
All contributions in spatial bin $j$ are summed up to yield
\begin{equation}
\left<  v^l_r v^m_\vartheta v^n_\varphi \right>^j_i \equiv 
\sum_{k \in {\cal J}^j} v^l_r v^m_\vartheta v^n_\varphi \, \frac{\Delta t^k_i}{T_i}.
\end{equation}

Thus for the whole library the velocity moments in spatial bin $j$ follow as
\begin{equation}
\label{libmoments}
\left< v^l_r v^m_\vartheta v^n_\varphi \right>^j = 
\frac{1}{\der L^j} \, \sum_i w_i \, \left< v^l_r v^m_\vartheta v^n_\varphi \right>^j_i.
\end{equation}

\subparagraph*{Projected kinematics.}
For the projected kinematics of the library we record the
normalized contribution to the 
kinematics $\mathrm{LOSVD}^{jk}_i$ at projected
position  $j$ and projected velocity $1 \le k \le N_\mathrm{vel}$ for every orbit.
Again for the whole library the LOSVD reads
\begin{equation}
\label{liblosvd}
\mathrm{LOSVD}^{jk} = \sum_i w_i \, \mathrm{LOSVD}^{jk}_i.
\end{equation}
By fitting a Gauss-Hermite series to the $\mathrm{LOSVD}^{jk}$ 
we obtain the Gauss-Hermite parameters (\citealt{Ger93}; \citealt{vdMF93})
\begin{equation}
\mathrm{GHP}^{jk} = \{ \gamma^{jk}, v^{jk}, \sigma^{jk}, H_3^{jk}, H_4^{jk} \}
\end{equation} 
of the LOSVD.

\subsection{Choice of orbits}
\label{orbits}

To obtain a reliable representation of phase-space it is important that any allowed
combination of the integrals of motion $(E,L_z,I_3)$ is represented to some degree of
approximation by an orbit in the library. The absence of some orbit family in the library 
might misleadingly emphasize certain dynamical configurations in the final fit.

\subparagraph*{Sampling $E$ and $L_z$.}
Richstone et al. (in preparation) adjust the orbit sampling in ($E,L_z$)-space according
to their spatial binning. From the requirement that every pair
of grid bins $r_i \le r_j$ in the equatorial plane should be connected 
by at least one equatorial orbit with $r_{\mathrm{peri}}=r_i$ and $r_{\mathrm{apo}}=r_j$
they uniquely derive a grid of orbital energies $E$ and z-angular momenta $L_z$.
We experimented with doubling the number of peri- and/or apocenters per radial
bin but found that the above described method yields a sufficiently dense sampling
of the ($E,L_z$)-plane.

\subparagraph*{Sampling $I_3$.}
It is common practice among the various existing Schwarzschild codes to sample $I_3$ by 
dropping orbits at given energy $E$ and angular momentum $L_z$ from the zero-velocity-curve 
(ZVC, defined by 
$E = L_z^2/(2 r^2 \cos^2 \vartheta) + \Phi(r,\vartheta)$). Richstone et al. (in preparation) 
use the intersections of the angular rays of the meridional grid with the ZVC as starting 
points. This sampling ensures that each sequence of orbits with common $E$ and $L_z$ contains 
at least one orbit that is roughly confined to the region between the equatorial plane and 
each angular ray of the meridional grid.

\begin{figure}
\includegraphics[width=60mm,angle=-90]{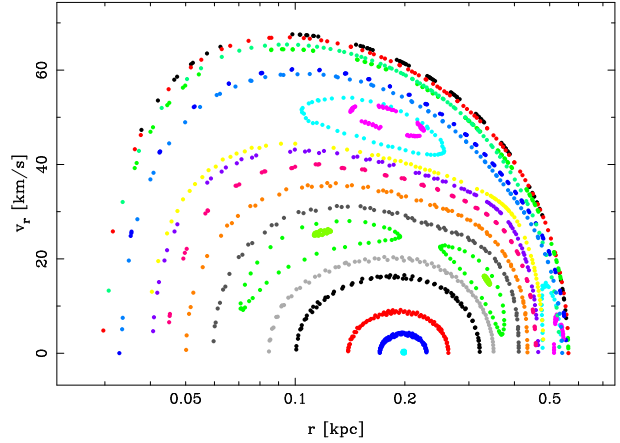}
\caption{Example of a surface of section for a flattened Hernquist model (details in 
the text). All orbits have been integrated for $N_\mathrm{SOS} = 80$ intersections with 
the SOS.}
\label{sos}
\end{figure}

If we consider only potentials symmetrical about the equatorial plane 
with $\der \Phi / \der z > 0$, then every orbit eventually crosses the equatorial plane 
and leaves a footstep in the surface of section (SOS) given by the radii $r$ and radial 
velocities $v_r$ of the upward equatorial crossings. Orbits respecting a third integral
show up in the SOS as nested invariant curves, sometimes with embedded resonances
(e.g. \citealt{binneytremaine}). Fig.~\ref{sos} shows an example of a SOS. The dots mark 
representative points of invariant curves obtained by numerically following orbits with 
common $E$ and $L_z$ in a flattened Hernquist potential with total mass
$M=10^{11} \mathrm{M}_{\sun}$, scaling radius $r_s=10 \, \mathrm{kpc}$ and a flattening
of $q=0.5$ (see Sect.~\ref{osmproj} below for further details). 

The SOS encompasses all 
available orbital shapes and a representative sampling of orbits should end up with the SOSs 
homogeneously filled with orbital imprints. Unfortunately, we are not aware of any simple 
relationship between the drop-point of an orbit on the ZVC and its corresponding appearance 
in the SOS, as long as $I_3$ is not known explicitly. In order to guarantee a representative 
collection of orbits in any case, we sample the orbits as follows.

In a first step we drop orbits from the (outer) intersections of the angular rays of our
spatial grid with the ZVC as described in Richstone et al. (in preparation). Then,
for any pair ($E,L_z$) included in the library we choose $N_L$ radii $r_l$,
$1 \le l \le N_L$ equally spaced
in $\log(r)$ on the equatorial plane between $r_{\mathrm{peri}}$ and $r_{\mathrm{apo}}$ 
of the equatorial radial orbit with energy $E$ and angular momentum $L_z$. 
We start with the smallest of these radii $r_l$ and launch an orbit $i$
from the equatorial plane with the maximal radial velocity
\begin{equation}
\label{vr start}
v_{r,i} = \sqrt{2\left(E-\Phi(r_l)\right)-\frac{L_z^2}{r^2_l}} 
\equiv v_\mathrm{max}(E,L_z,r_l).
\end{equation}
For the subsequent orbits $i'$ we stepwise decrease $v_{r,i'}$ by
$\Delta v_{r,i'}$ (see equation~(\ref{vstep}) below) until we reach $v_{r,i'} = 0$ and
pass over to the next radius $r_{l+1}$.

With ($E,L_z$) and ($r_l,v_{r,i}$) fixed the orbital $v_{\vartheta,i}$ is determined by
\begin{equation}
\label{vthe start}
v_{\vartheta,i} (E,L_z,r_l,v_{r,i}) = \sqrt{2\left(E-\Phi(r_l) \right)-v_{r,i}^2-
\frac{L_z^2}{r_l^2}}.
\end{equation}
When $v_{r,i} = 0$, then $v_{\vartheta,i}(E,L_z,r_l,v_{r,i}) = v_\mathrm{max}(E,L_z,r_l)$.
For each velocity pair we launch an orbit from the equatorial plane
at the actual $r_l$ with the actual velocities $v_{r,i}$ and $v_{\vartheta,i}$. 
This procedure is repeated for each of the $N_L$ radii.
If, at a specific launch position, we find an imprint in the SOS of a previously 
integrated orbit which differs from the current launch position by less than 10 per cent 
in radius and radial velocity, then we regard this part of the SOS as already sampled and 
discard the orbit.

The velocity stepsize $\Delta v_{r,i}$ is set as
\begin{equation}
\label{vstep}
\Delta v_{r,i} = \min \left \{ \Delta v_\mathrm{LOSVD} \, , \, 
\xi \ v_{m,i-1} \right \},
\end{equation}
where $\Delta v_\mathrm{LOSVD}$ is the width of the LOSVD-bins (cf. equation~(\ref{delta LOSVD}))
and 
\begin{equation}
v_{m,i} = \max_{1\le s \le N_\mathrm{sos}} \left \{
v^s_i : (r^s_i,v^s_i) \in \mathrm{SOS} \right \}.
\end{equation}
Here $\mathrm{SOS}$ denotes the set of the $N_\mathrm{sos}$ orbital imprints in the SOS
and $i$ is the index of the actual orbit. We usually take $\xi = 1/3.5$. 
Trying different values for $N_L$ we
found $N_L = 30$ sufficient to yield a dense filling of the SOS with approximately
one invariant curve crossing the $r$-axis of the SOS in each of the equatorial meridional 
grid bins. The corresponding orbits have large $\vartheta$-motion and need to be included
in the library to avoid a radially biased collection of orbits.

The velocity stepsize is largest for the radial orbits
and gradually decreases when the SOS is filled with orbits (note that $v_{m,i-1}$ 
is the maximum of the radial velocities in the SOS of the ``precursor'' orbit $i-1$). 
For the shell orbits, the stepsize becomes smallest. The adjustment of the stepsize in 
each step
ensures that we sample the more radial orbits with a resolution that corresponds at least
to the width of the LOSVD bins and that the sampling is refined for the shell orbits. 

After the above described sampling is done, we measure the maximum 
$f_s$ of all $r_{\mathrm{min},i}/r_{\mathrm{max},i}$, with
\begin{equation}
r_{\mathrm{min},i} = \min_{1\le s \le N_\mathrm{sos}} \left \{
r^s_i : (r^s_i,v^s_i) \in \mathrm{SOS} \right \}
\end{equation}
and $r_{\mathrm{max},i}$ defined analogously. To ensure that the sequence contains all
orbits up to (approximately) the thin shell orbit, we complete the library if necessary by
launching orbits from the equatorial plane with $v_r = 0$ at
\begin{equation}
r = \frac{3 \, r_{\mathrm{min},i'} + r_{\mathrm{max},i'}}{4},
\end{equation}
where $f_s = r_{\mathrm{min},i'}/r_{\mathrm{max},i'}$, until $f_s>0.9$.

Fig.~\ref{sos} illustrates  the dense coverage of the SOS with invariant curves after 
all orbits are integrated for a flattened Hernquist potential.

\subsection{Use of the library}

If the relative contribution of each orbit to the whole library, the orbital weight
$w_i$, is specified, then according to equations (\ref{def dL}), (\ref{libmoments}) and 
(\ref{liblosvd}) the library provides a specific model including the LOSVDs, 
internal density distribution, internal velocity moments and so on of this
particular orbit superposition.

If the library is constructed to test whether or not a given trial mass distribution
leads to a consistent model of an observed galaxy, then the model, in particular the
LOSVDs, have to be compared to the observations. If the comparison turns out not to yield
a satisfactory fit, then the weights can either be recalculated (see Sec.~\ref{libfit} for
details) or the actual mass distribution has to be rejected. If on the other hand, the
fit shows that the actual set of weights seems to be a valid model of the galaxy, then
one can reconstruct the internal velocity structure and DF from the $w_i$'s.

Conversely, if one has a DF at hand and wants to calculate e.g. its projected
kinematics without going through the appropriate integrals, one can assign the orbital
weights according to the DF (see Sec.~\ref{projdf}) without any fitting procedure and
analyse the output of the library. This can be usefull to study systematically the
projected properties of general axisymmetric distribution functions depending on all
three integrals $(E,L_z,I_3)$.

In the following we will make use of both applications with the goal to investigate
the accuracy of our orbit libraries.


\section{Orbital weights and phase-space densities}
\label{weights and densities}

In order to reconstruct the DF from the library or to calculate spatial profiles
of internal or projected properties of some given DF, one needs to convert orbital 
weights into phase-space densities and vice versa. This section summarizes the connection 
between orbital weights and orbital phase-space densities under the regime of Jeans' theorem. 

\subsection{Phase-space densities of orbits}
\label{phasedens}
Consider a system in which the orbits respect $n$ integrals of motion
$I_1,\ldots,I_n$.
Because the phase-space density of stationary systems is constant along individual orbits
(Jeans theorem) the phase-space density along orbit $i$ is given as the orbital weight
$w_i$ divided by the phase-space volume $V_i$. More formally, let ${\cal I}$ denote the 
$n$-dimensional set of orbital integrals $(I_1,\ldots,I_n)$, let ${\cal V}$ denote the 
6-dimensional phase-space, ${\cal P}({\cal V})$ its power set and let
$\xi : {\cal I} \rightarrow {\cal P}({\cal V})$ map a $n$-tuple of orbital integrals
$(I_1,\ldots,I_n) \in {\cal I}$ on to the hypersurface 
$\xi(I_1,\ldots,I_n) \subseteq {\cal V}$
in phase-space covered by the corresponding orbit,
\begin{equation}
\xi(I_1,\ldots,I_n) \equiv \{ p \in {\cal V} : I_1(p)=I_1,\ldots,I_n(p)=I_n \}.
\end{equation} 
With ${\cal U}_i \subseteq {\cal I}$ being the small cell in
integral space represented by the orbit $i$,
\begin{eqnarray}
\lefteqn{{\cal U}_i \equiv \{ (I_1,\ldots,I_n) \in {\cal I} : I_1 \in 
[I_{1,i} - \Delta I_{1,i},I_{1,i} + \Delta I_{1,i}],} \nonumber \\
& & \ldots, I_n \in [I_{n,i} - \Delta I_{n,i},I_{n,i} + \Delta I_{n,i}] \}
\end{eqnarray}
we define the characteristic function
\begin{equation}
\chi_i\equiv\left\{ 
\begin{array}{r@{\quad:\quad}l}
1 & (r,\vartheta,\varphi,v_r,v_{\vartheta},v_{\varphi}) \in 
{\cal O}_i \\
0 & (r,\vartheta,\varphi,v_r,v_{\vartheta},v_{\varphi}) \notin 
{\cal O}_i
\end{array}
\right.
\end{equation}
of the image set 
\begin{equation}
{\cal O}_i \equiv \bigcup_{{\cal W} \in \xi({\cal U}_i)} {\cal W}
\end{equation}
of ${\cal U}_i$ in phase-space.
The volume of the phase-space region represented by orbit $i$
then follows as
\begin{equation}
\label{phvoldef}
V_i = \int \chi_i \, \der^3r \, \der^3v
\end{equation}
and accordingly the phase-space density along the orbit reads
\begin{equation}
\label{orbdf}
f_i \equiv \frac{w_i}{V_i}.
\end{equation}

\subsection{Orbital weights from DFs}
\label{orbweights}
If we reverse the application of equation~(\ref{orbdf}), and
assign the orbital weights according to some given DF $f$,
\begin{equation}
\label{weights}
w_i = f_i \, V_i,
\end{equation}
with $f_i \equiv f (I_{1,i},\ldots,I_{n,i})$ now being the DF $f$ evaluated at the orbit's
position in integral space,
then the DF $f_\mathrm{lib}$ of the entire library, which consists of the combined
contributions of all orbits
\begin{equation}
\label{libdf}
f_{\mathrm{lib}} = \sum_i f_i \chi_i
\end{equation}
will be the mapped version of $f$ on to the library.
Equation~(\ref{weights}) together with equations (\ref{def dL}), (\ref{libmoments}) and 
(\ref{liblosvd}) can be used to calculate the LOSVDs, internal velocity profiles and 
density distribution of any axisymmetric DF with known potential.


\section{Orbital phase-volumes}
\label{phasevols}
\subparagraph*{Two degrees of freedom.}
\citet{binney85} have shown, that for autonomous Hamiltonian systems with two degrees of freedom
the phase-volume of any orbit can be derived from the SOS by integrating 
the times between successive orbital visits of the SOS
\begin{equation}
V \approx \Delta E \, \int_\mathrm{SOS} T(r,v_r) \, {\mathrm d}r \, {\mathrm d}v_r,
\end{equation}
where $T(r,v_r)$ is the time the orbit needs from $(r,v_r)$ to the next 
intersection with the SOS, and $\Delta E$ defines a small but finite cell around
the orbit's actual energy $E$ characterizing the hypersurface in
phase-space represented by the orbit.

\subparagraph*{Axisymmetric case.}
Richstone et al. (in preparation) carry over this result to axisymmetric systems
and approximate the phase-volumes as
\begin{equation}
\label{phase}
V \approx \Delta L_z \, \Delta E \,
 \int_\mathrm{SOS} T(r,v_r) \, {\mathrm d}r \, {\mathrm d}v_r.
\end{equation}
Here $\Delta L_z$ and  $\Delta E$ stand for the range of energies and
angular momenta represented by the orbit under consideration. Equation~(\ref{phase})
is valid independent from the orbit beeing regular or chaotic.

\subparagraph*{Calculating the SOS-integral.}
In what follows we describe our novel implementation of equation~(\ref{phase}) 
that improves on the method of Richstone et al. (in preparation) to deliver higher 
precision phase-space volumes.

For all orbits in a sequence with common energy $E$ and angular momentum
$L_z$ we obtain a representative sample ${\cal S}$ of the SOS by storing
$N_\mathrm{sos}$ imprints of each orbit in the SOS given by the radial
positions and velocities\footnote{To reduce the computational 
effort we take the absolute value
of the radial velocities, thereby exploiting the symmetry of the SOS with
respect to the $r$-axis in our application.} at the
times $t_i^{k(s)}$ of the orbital equatorial crossings,
\begin{eqnarray}
{\cal S} \equiv \big \{
(r^s_i,v^s_i):
r^s_i \equiv r(t_i^{k(s)}),
v^s_i \equiv |v_r(t_i^{k(s)}) |, \nonumber \\
E_i=E, L_{z,i}=L_z, 
1 \le s \le N_\mathrm{sos}
\big \}.
\end{eqnarray}
Typically, we integrate each orbit up to $N_\mathrm{sos}=80$ intersections with the SOS
and choose $N^\prime_\mathrm{sos}=60$ points for the 
calculation of the phase-volumes randomly out of the whole set of intersections. 
We also store the time intervals 
\begin{equation}
t(r^s_i,v^s_i) \equiv t^{k(s+1)}_i - t^{k(s)}_i
\end{equation}
between two successive intersections. 

Inspection of Fig.~\ref{sos} shows that only a tessellation approach can
numerically integrate equation~(\ref{phase}) in the general case including
regular, resonant and chaotic orbits. To this purpose we decided to perform
a Voronoi tessellation of ${\cal S}$ using the software of \citet{shew}.
This tessellation uniquely allocates a polygon to each element of ${\cal S}$.
The edges of the polygon are located on the perpendicular bisections of pairs containing
the element under consideration and one of its neighbours and are
equidistant to the actual pair and a third element. For almost all elements
the polygons are closed and encompass an area containing the actual element and
all points that are closer to it than to any other element. 
The areas enclosed by the polygons completely cover the space between the elements and
therefore characterize the fractional area inside the SOS occupied by each orbit.

\begin{figure}
\includegraphics[width=60mm,angle=-90]{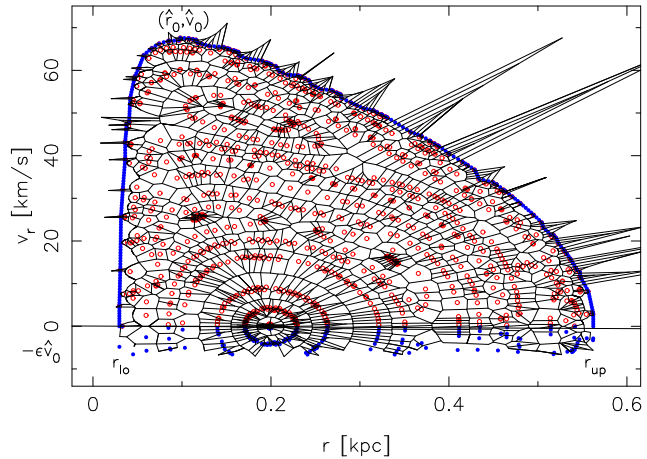}
\caption{A Voronoi tessellation of the SOS of Fig.~\ref{sos}. Open circles mark individual 
intersections of orbits with the SOS, solid dots are points added to get the Voronoi cells 
well behaved at the boundaries.}
\label{voronoisos}
\end{figure}

Fig.~\ref{voronoisos} shows the same SOS as Fig.~\ref{sos}. The open circles display $r$ and
$v_r$ at the orbital equatorial crossings. The thin lines around these circles mark
the Voronoi cells, e.g. the polygons allocated to the elements of ${\cal S}$ and
boundary points (see below).

With $\Delta A^s_i$ beeing the surface area enclosed
by the polygon around \mbox{$(r^s_i,v^s_i) \in {\cal S}$} the integral expression in the
phase-volume of orbit $i$ (cf. equation~(\ref{phase})) can be approximated\footnote{Note 
that the Poincar\'e map of the SOS on to itself is area-preserving and one would like to
have $\Delta A^s_i$ independent of $s$. The Voronoi
tessellation however yields only approximately constant $\Delta A^s_i$. Nevertheless,  
as Sec.~\ref{projdf} shows, the resulting phase-volumes are of high
accuracy.} as
\begin{equation}
\label{voronoi}
\int_\mathrm{SOS} T(r,v_r) \, {\mathrm d}r \, {\mathrm d}v_r
\approx \sum_s t(r^s_i,v^s_i) \ \Delta A^s_i.
\end{equation}

At the boundary of the distribution of sampled points, there may not be enough
neighbours around a given element of ${\cal S}$ to close its polygon.
In order to ensure that every Voronoi polygon is closed and confined to an area enclosed by
the ZVC of the SOS (given by $v_r$ of equation~(\ref{vr start}))
we construct an envelope around the distribution of sampled orbital intersections. In
Fig.~\ref{voronoisos} the points of the envelope are marked by the solid dots.
They are constructed as follows.

In a first step, we determine the maximum $\hat{v}_0$ of 
radial velocities in ${\cal S}$
\begin{equation}
\hat{v}_0 \equiv \mathrm{max} \left \{ 
v_r : (r,v_r) \in {\cal S} \right \}.
\end{equation}
To ensure that no Voronoi cell exceeds below the axis $v_r = 0$,
all imprints in the SOS with $v^s_i \le \epsilon \, \hat{v}_0$ are mirrored about
the axis $v_r = 0$ (typically $\epsilon = 0.1$).
For the rest of the SOS we construct an envelope in an iterative loop 
starting from
\begin{equation}
(\hat{r}_0,\hat{v}_0) \equiv (r,v_r) \in {\cal S}, \; v_r=\hat{v}_0.
\end{equation}
In each iteration $n+1$ we search for $(\hat{r}_{n+1},\hat{v}_{n+1}) \in {\cal S}$ obeying
\begin{equation}
\hat{v}_{n+1} = \mathrm{max} \left \{ v_r : (r,v_r) \in {\cal S}, \; r > \hat{r}_n \right \}.
\end{equation}
We compose the envelope out of points densely sampled from the line segment 
connecting ($r'_n,v'_n$) and ($r'_{n+1},v'_{n+1}$), 
where the prime indicates that the coordinates
are slightly shifted outwards, i.e. $r'_n = (1+\delta) \hat{r}_n$ and 
$v'_n = (1+\delta) \hat{v}_n$ with typical $\delta = 0.01$. 
The loop eventually stops after $N$ iterations at
$\hat{r}_N = r_\mathrm{up}$ and is followed by an analogous procedure running from 
$\hat{r}_0$ to $r_\mathrm{lo}$ to complete the envelope along the left part of the SOS.
The points on the envelope are used as additional seeds for the Voronoi tessellation.

As Fig.~\ref{voronoisos} shows,
the apposition of the boundary points as described above ensures that all
orbital Voronoi cells are closed and confined to an area roughly bounded by the outermost
invariant curve of the SOS. The definition of the boundary is purely geometrical and
insensitive to numerical uncertainties in the orbit integration. 
The spiky cells along the upper boundary belong to seeds of the envelope and do not 
affect the calculation of the orbital phase-volumes.

The Voronoi tessellation used to approximate the integral expression in
equation~(\ref{phase}) via equation~(\ref{voronoi}) defines a robust method
to calculate the relative phase-volume of any orbit inside a particular sequence of orbits
with common $E$ and $L_z$, including resonances and chaotic orbits.
The areas assigned to the individual orbital imprints in the
SOS completely fill the area below the ZVC of the SOS. Thus, a cruder sampling of
the SOS is compensated by larger individual orbital phase-volumes.
In the limit of an infinitely dense sampling,
the assigned ``phase-space-weights'' obtained by the tessellation approach
single-orbit phase-volumes.

\subparagraph*{Calculating $\Delta E \, \Delta L_z$.}
For a complete determination of the phase-volumes we also need the relative
contribution of a whole sequence of orbits with common ($E,L_z$) as compared 
to other sequences with different energies and angular momenta. These are described by the
factors $\Delta L_z \, \Delta E$ of the orbital phase-volumes (cf. equation~(\ref{phase})). 
They are in fact equal for all orbits in the same
sequence and need to be calculated only once for each sequence. 

\begin{figure}
\includegraphics[width=60mm,angle=-90]{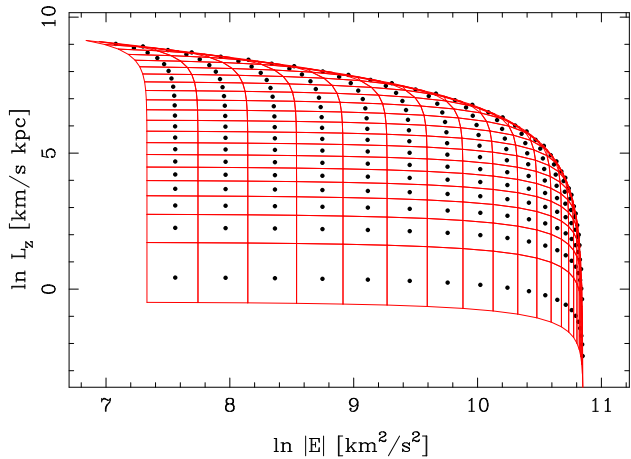}
\caption{Typical distribution of sampled energies $E$ and angular momenta $L_z$ of an orbit
library. The thin red lines show the boundaries of small cells assigned to each sequence.
Their surface area is taken to estimate $\Delta E \, \Delta L_z$. The potential equals
that of Figs.~\ref{sos} and \ref{voronoisos}.}
\label{elz}
\end{figure}

Fig.~\ref{elz} shows an example of the ($E,L_z$)-plane of a typical library. 
The dots display the grid of sampled orbital energies and angular momenta. 
Each dot represents a 
sequence of orbits with common $E$ and $L_z$ but different $I_3$. 
To calculate $\Delta L_z \, \Delta E$ for a particular 
sequence, we construct a quadrangle around the sequence's ($E,L_z$) and estimate the product 
$\Delta L_z \, \Delta E$ as the surface area enclosed by this quadrangle. The thin
lines in Fig.~\ref{elz} show the boundaries of these quadrangles, which are constructed
as follows.

As described in Sec.~\ref{spatial grid}
the grid in ($E,L_z$)-space is derived from the requirement that for every pair
$r_i < r_j$ of equatorial grid bins, the library contains at least one 
equatorial orbit with $r_\mathrm{peri}=r_i$ and $r_\mathrm{apo}=r_j$.
Consider now a sequence of orbits
with ($E_\mathrm{seq},L_{z,\mathrm{seq}}$) and corresponding
$r_\mathrm{peri,seq}$ and $r_\mathrm{apo,seq}$ of the equatorial orbit. In ($E,L_z$)-space
all sequences inside the boundary of the sampled area are surrounded
by four other sequences having both their peri- and their apocenter in 
adjacent spatial bins. Let 
$r_{\mathrm{peri},j}$ and $r_{\mathrm{apo},j}$, $1 \le j \le 4$, 
denote the corresponding peri- and apocenters of the equatorial orbits of these sequences.
We construct a quadrangle around the sequence ($E_\mathrm{seq},L_{z,\mathrm{seq}}$)
by connecting the energies and angular momenta 
of four ficticious orbit sequences characterised by the
pericenter $\hat{r}_{\mathrm{peri},j}$ and apocenter $\hat{r}_{\mathrm{apo},j}$ of 
their equatorial orbits
\begin{equation}
\hat{r}_{\mathrm{peri},j} = \frac{1}{2} (r_{\mathrm{peri},j}+r_\mathrm{peri,seq})
\end{equation}
and
\begin{equation}
\hat{r}_{\mathrm{apo},j} = \frac{1}{2} (r_{\mathrm{apo},j}+r_\mathrm{apo,seq}).
\end{equation}

The sequences with the largest apocenters and the smallest pericenters, respectively, 
are surrounded by less than four sequences having both their peri- and their 
apocenter in adjacent spatial bins. For these sequences we calculate the edges of 
the quadrangle as if there were further sequences around, whose energies and
angular momenta follow from our spatial grid at smaller
radii than $r_\mathrm{min}$ and larger radii than $r_\mathrm{max}$.

Sequences with $r_\mathrm{peri,seq} \approx r_\mathrm{apo,seq}$ (lying on the
upper boundary of the sampled area in Fig.~\ref{elz} and usually containing only
one, approximately circular, orbit) are also not surrounded by four sequences
as described above. For these sequences we take 
($E_\mathrm{seq},L_{z,\mathrm{seq}}$) of the actual sequence as the upper right edge
of the quadrangle.

As can be seen in Fig.~\ref{elz} the quadrangles around the sequences' energies and angular
momenta completely cover the sampled part of the allowed area in ($E,L_z$)-space below
the curve $L_z(E) = L_{z,\mathrm{circ}}$. They give a reasonable measure of 
the fractional area in ($E,L_z$)-space, occupied by each orbit sequence.

\section{Mapping distribution functions on to the library}
\label{projdf}

In this Section we describe how to use the phase-volumes from the previous Section
to calculate internal and projected properties of stationary DFs using an orbit library.
To this end, starting with a density profile $\rho$ and a stationary
distribution function $f_{\rho}$ connected to $\rho$ via $\rho = \int f_{\rho} \, \der^3 v$,
a library is constructed as described in Sec.~\ref{library}. Instead of fitting
the library to the kinematics of $f_{\rho}$,
\begin{equation}
\label{losvdf} 
\mathrm{LOSVD}_f(v_{los}) = \frac{1}{\rho} \int f_{\rho} \, \der^2 v_{\bot}
\end{equation}
we {\it assign} an appropriate weight to each orbit such that the superposition 
of all orbits represents $f_{\rho}$ (see Sec.~\ref{orbweights} above). 
We then compare the internal density distribution $\rho_{\mathrm{lib}}$ and 
the anisotropy profile $\beta_\mathrm{lib}$, as well as the projected kinematics 
$\mathrm{GHP}_{\mathrm{lib}}$
obtained from the library with the same properties $\rho$, $\beta$ and
$\mathrm{GHP}$ directly calculated from the DF (see Sec.~\ref{osmproj} - \ref{plumproj}). 
Thereby we can check to which
accuracy the orbit library reproduces a {\it given} dynamical system.

\subsection{Spherical $\gamma$-models}
\label{osmproj}
As a first reference case, we explore spherical $\gamma$-models.

\subparagraph*{Properties of the input model.}
The stellar body of the reference model is constructed from
$\gamma$-models \citep{D93} with density
\begin{equation}
\label{gammasph}
\rho_\gamma(r) = \frac{M}{4 \pi} \frac{r_s(3-\gamma)}{r^{\gamma}(r_s+r)^{4-\gamma}}.
\end{equation} 
They approximate the de Vaucouleurs law of ellipticals quite well for $\gamma \in [1,2]$. 
The DF is assumed to be of the Osipkov-Merritt 
type $f_\mathrm{OM} = f_\mathrm{OM} (E-L^2/2r_a^2)$  
(\citealt{O79}; \citealt{M85a}; \citealt{M85b}). The
corresponding systems are isotropic at $r \ll r_a$ and radially anisotropic at $r \gg r_a$:
\begin{equation}
\label{ombeta}
\beta \equiv 1 - \frac{v_\vartheta^2 + v_\phi^2}{2 \, v_r^2} = \frac{r^2}{r^2 + r_a^2}.
\end{equation}
We tested various combinations of the parameters ($\gamma,r_s,r_a$). However,
since the conclusions drawn from the comparisons do not depend strongly on $\gamma$,
the following contains only a discussion of the results for the Hernquist model 
($\gamma = 1$), where the DF can be written in terms of elementary functions and 
reads \citep{hernquist}
\begin{eqnarray}
\label{dfhern}
f(E,L) \propto \frac{1}{8 (1-q^2)^{5/2}} - 3 \arcsin q + (1 - 2 q^2) \nonumber \\ 
\left( q \sqrt{1-q^2} (8 q^4 - 8 q^2 - 3) + \frac{r_s^2}{r_a^2} q 
\right)
\end{eqnarray}
with $q= \sqrt{r_s(E-L^2/2r_a^2)/G M}$.

\begin{figure}
\includegraphics[width=84mm,angle=0]{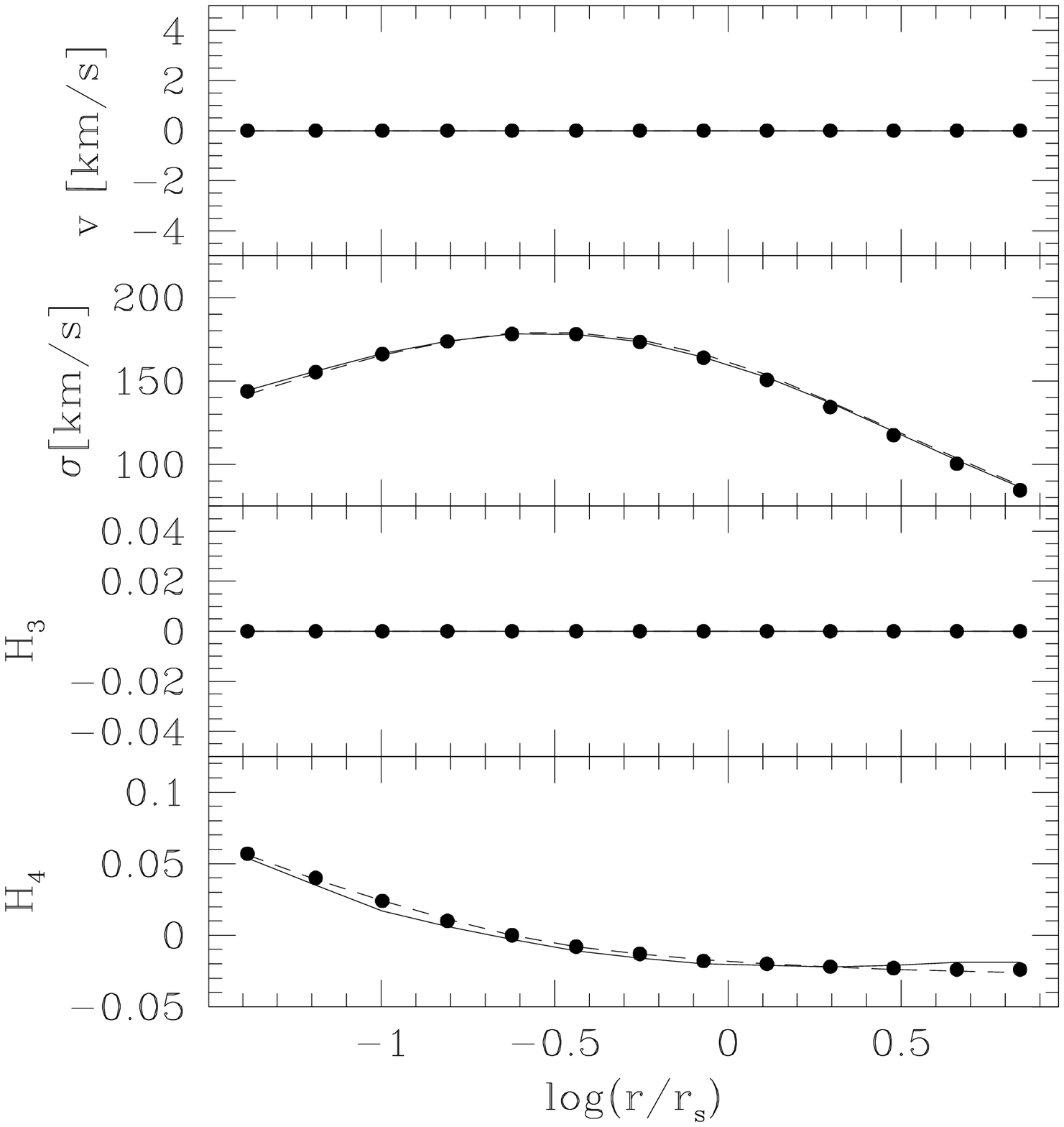}
\includegraphics[width=84mm,angle=0]{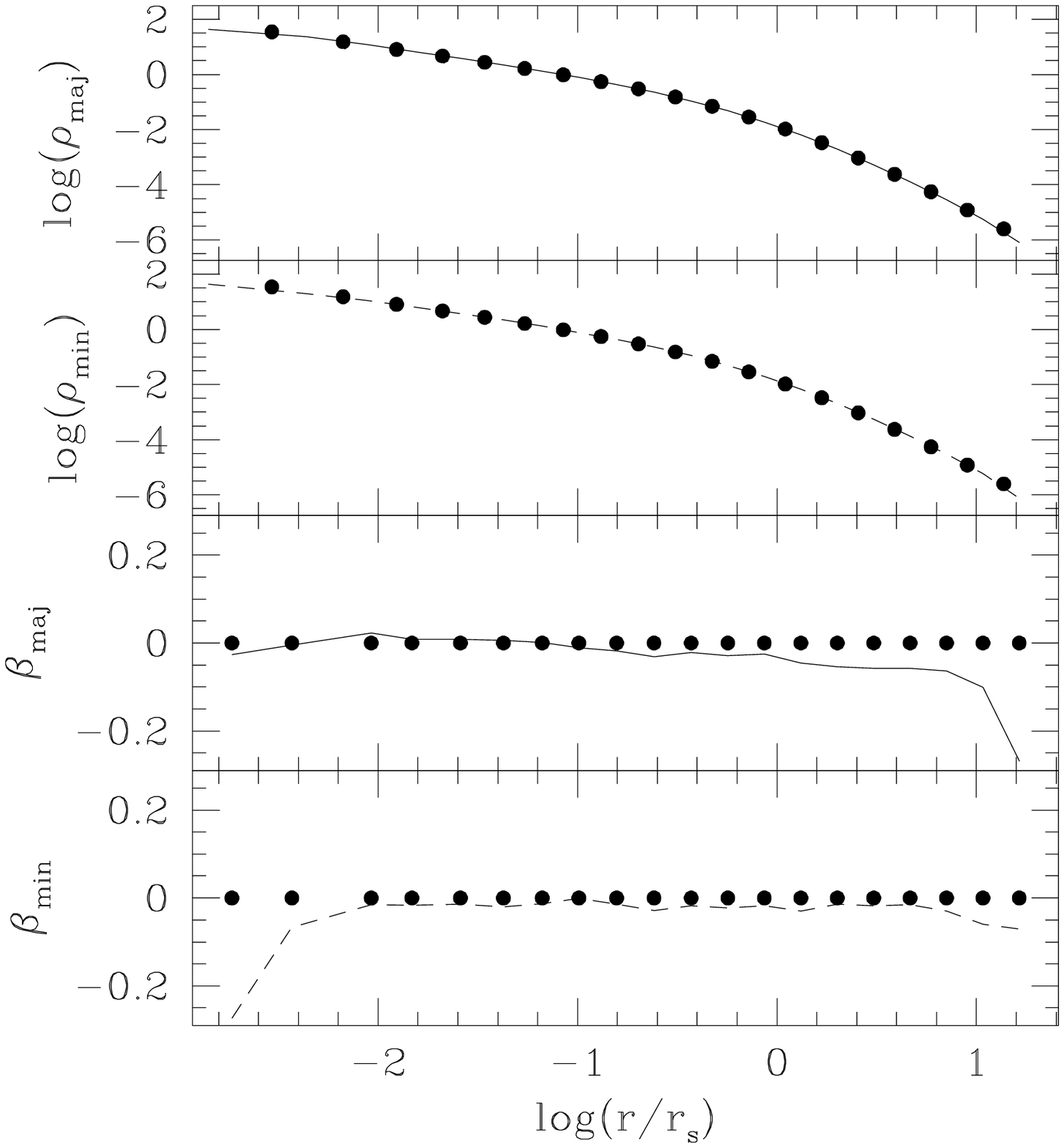}
\caption{Comparison of a Hernquist model (big dots) and a library with weights directly
derived from the spherical, isotropic Hernquist DF (lines). The upper panel shows 
the projected kinematics along the major axis (solid line) and minor axis (dashed
line). The lower panel shows the density distribution (upper two rows, 
$[\rho] = \mathrm{M}_{\sun}/\mathrm{pc}^3$) and the anisotropy parameter (lower two rows) for the minor and 
major axis, respectively.}
\label{proj.hern.iso}
\end{figure}

\subparagraph*{Comparison of model and library.}
Fig.~\ref{proj.hern.iso} shows the GHPs, density and anisotropy profiles 
of a library with $\approx 2 \times 8800$ orbits, extending from 
$\approx 5 \times 10^{-4} \, r_s$ to $\approx 28 \, r_s$. For this library we have
used a closed meshed sampling containing two different pericenters for each radial
bin. The weights for the orbits were directly derived from 
equation~(\ref{weights}) and the Hernquist DF of equation~(\ref{dfhern}) with 
$r_s = 10.5 \, \mathrm{kpc}$, a total mass of $M = 7.5 \times 10^{11} \mathrm{M}_{\sun}$ and 
$r_a = \infty$ (isotropic model). 
The big dots show the expected kinematics, density and anisotropy of the Hernquist
model. The GHPs were obtained by first calculating the LOSVDs 
at the position of the corresponding spatial bin from the DF using the 
method described in \citet{carollo95} and then fitting a GH-series to the LOSVDs.
For the density distribution and anisotropy we used equations (\ref{gammasph}) and 
(\ref{ombeta}), respectively. 

As the figure shows, the library is able to reproduce the GHP and internal density 
distribution of the model to a high degree of accuracy. The mean fractional difference
in $\sigma$ is below $\Delta \sigma <1$ per cent and the mean difference in $H_4$ is below 
$\Delta H_4 < 0.01$. The largest deviations between
model and library occur in the anisotropy profile with $\mathrm{rms}(\beta) = 0.06$
(taken over a whole angular ray). The individual differences however are smaller than
$\Delta \beta = 0.1$ over almost the whole spatial range covered by the library. 
Near the inner
and outer boundary of the library the orbit sampling becomes incomplete with mostly
radial orbits coming either from outside the outer boundary or from inside the inner
boundary are missing. Consequently, the library shows a decrease in radial velocity 
dispersion ($\beta < 0$) as compared to the expectations of the isotropic reference model.

\begin{figure}
\includegraphics[width=84mm,angle=0]{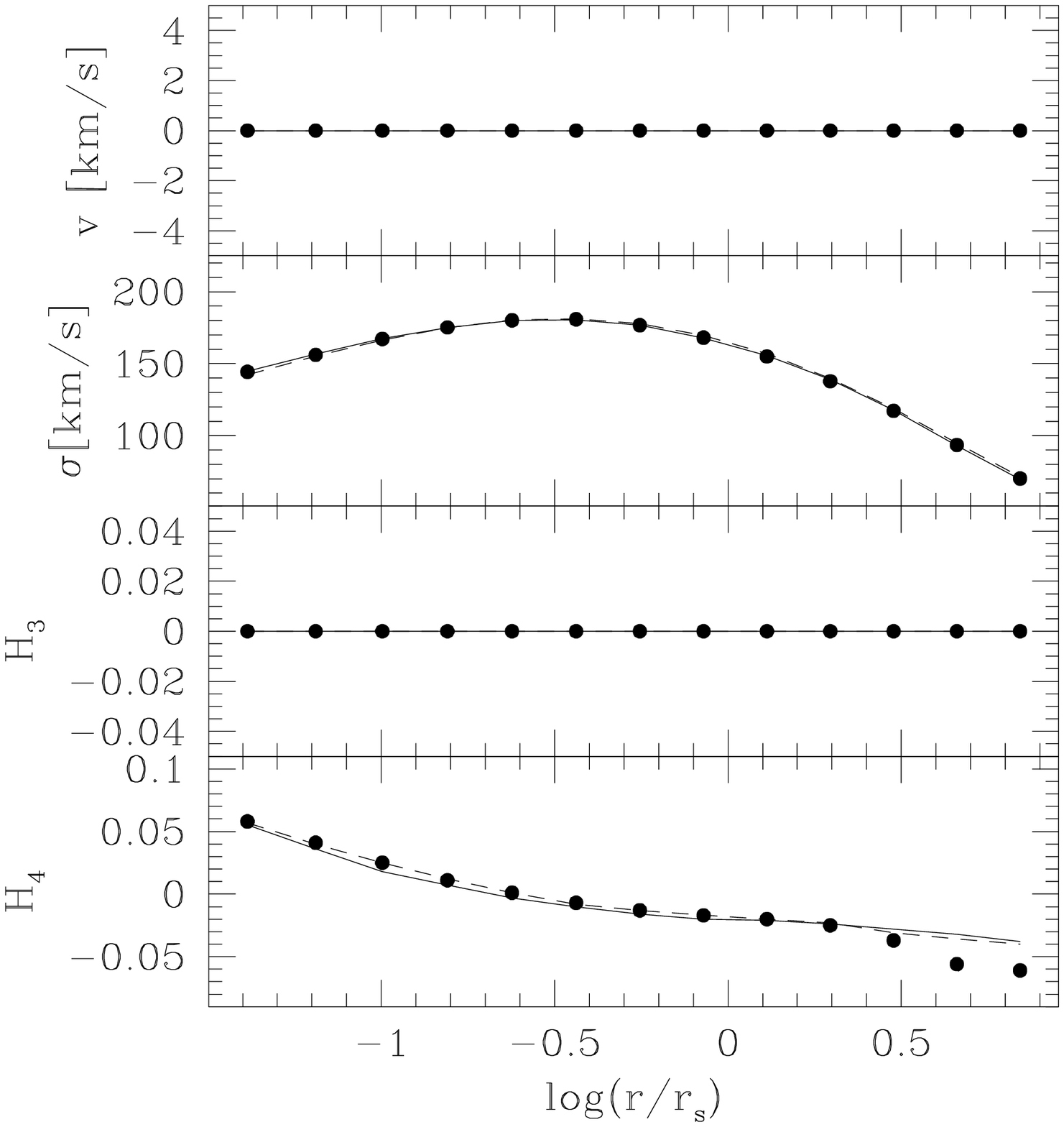}
\includegraphics[width=84mm,angle=0]{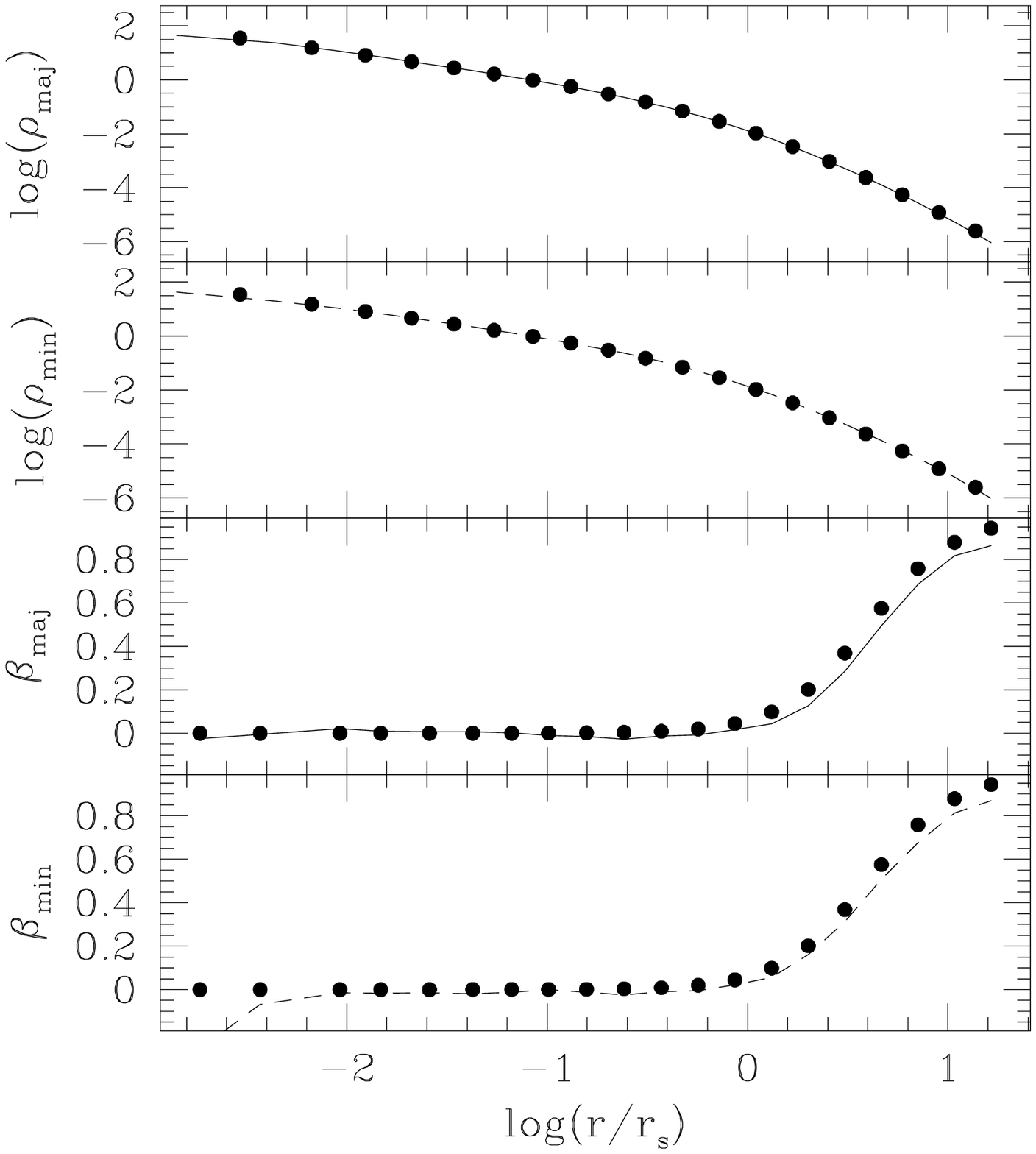}
\caption{Same as Fig.~\ref{proj.hern.iso} but for an anisotropic Hernquist
model with $r_a = 4 \, r_s$.}
\label{proj.hern.aniso}
\end{figure}

Fig.~\ref{proj.hern.aniso} shows the same as Fig.~\ref{proj.hern.iso} but for
an anisotropic Hernquist model with $r_a = 4 \, r_s$. It confirms the
results from the isotropic model. The offset in the $H_4$-profiles
at large radii is due to errors in the GH-fit. At these radii the
resolution of our LOSVD-bins is too low to give reliable GHPs. However the match of the
individual LOSVDs itself is as good as at smaller radii.

Again the largest deviations show up in the $\beta$-profiles, with a mean
$\mathrm{rms} (\beta_\mathrm{hern} - \beta_\mathrm{lib}) = 0.03$. As in the isotropic case
the differences between model and library increase when approaching the edges of the
library, where the radial velocity dispersion of the library is systematically lower than
expected.

\subsection{Flattened Plummer model}
\label{plumproj}
We now go one step further and use a {\it flattened}
test object, namely the flattened Plummer model of \citet{LB62} 
(normalized such that in the spherical limit $M$ defines
the total mass of the model)
\begin{eqnarray}
\label{plummer}
\rho_\mathrm{pl}(r,\vartheta) = \frac{M \lambda^{-9/4}}{4 \pi} 
[ 
\left( 3a^2-2b^2 \right) \left( r^2+a^2 \right)^2+ \nonumber\\
\left( 4a^2-b^2 \right) b^2r^2 \cos^2 (\vartheta)
],
\end{eqnarray}
$\lambda=\left(r^2 + a^2 \right)^2 - 2 b^2 r^2 \cos^2 (\vartheta)$. The parameters
$a$ and $b$ describe the extension of the core and the flattening.
The part of the distribution function, which is even in $L_z$ is 
\begin{eqnarray}
\label{plummerdf}
f_\mathrm{pl}(E,L_z) = \frac{\sqrt{2}}{4 \pi^{3/2}} \left(
\frac{\Gamma(6)}{\Gamma(\frac{9}{2})} D E^{\frac{7}{2}} + 
\frac{\Gamma(10)}{\Gamma(\frac{15}{2})} C L_z^2 E^{\frac{13}{2}}
\right),
\end{eqnarray}
$C \propto (3a^2 - 2 b^2)$ and $D \propto 5 b^2 (2 a^2 - b^2)$. The
Plummer models do not rotate as long as the uneven part of $f_\mathrm{pl}$ vanishes and 
prograde and retrograde orbits exactly balance each other.

\begin{figure}
\includegraphics[width=84mm,angle=0]{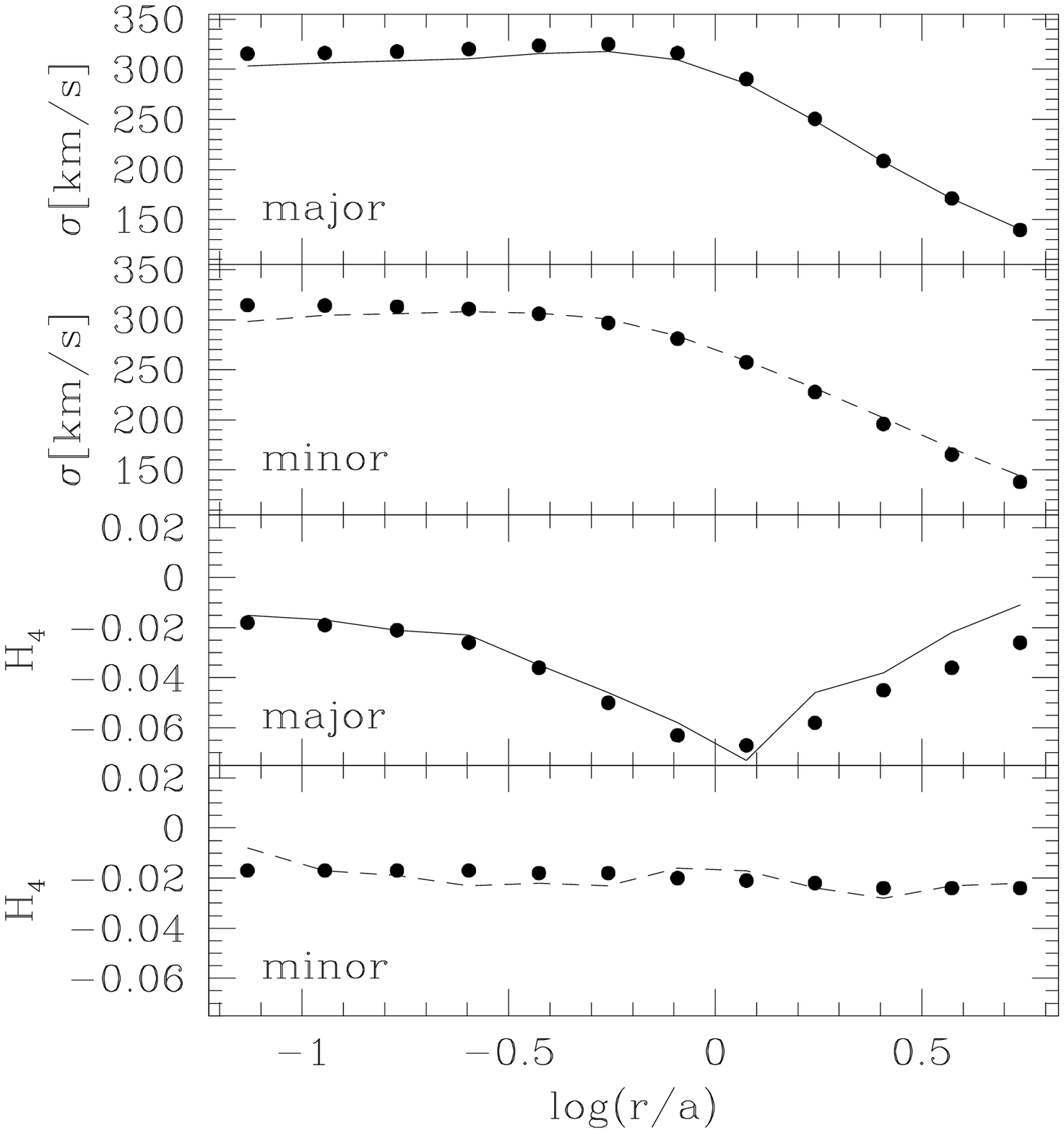}
\includegraphics[width=84mm,angle=0]{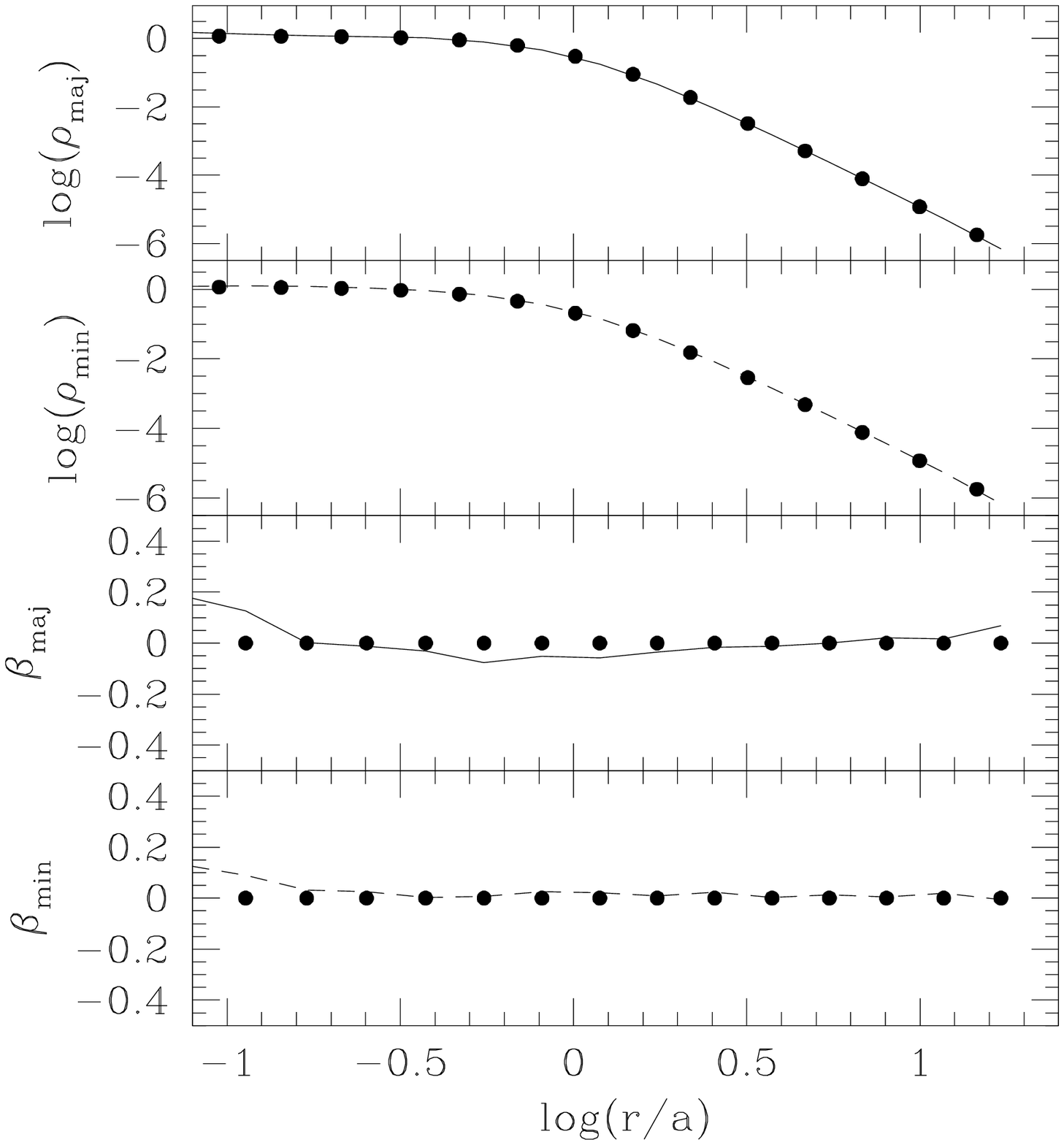}
\caption{Comparison of a flattened Plummer model (big dots) and a library 
with weights directly derived from the DF (lines). Upper panel: 
projected kinematics along major axis (solid lines) and minor axis (dashed
lines). Only moments independent from the uneven part of the DF are shown. Lower 
panel: density distribution (upper two rows, 
$[\rho] = \mathrm{M}_{\sun} \, \mathrm{pc}^{-3}$) and the
anisotropy parameter (lower two rows) for the two axes.}
\label{proj.plummer}
\end{figure}

\subparagraph*{Comparison of model and library.}
Fig.~\ref{proj.plummer} shows a flattened Plummer model
with $a = 5.0 \, \mathrm{kpc}$, $b = a/2$ and $M=7.5 \times 10^{11} \mathrm{M}_{\sun}$ 
(big dots)
and profiles obtained from a library with $\approx 2 \times 4400$ orbits,
extending from $\approx 10^{-3} \, a$ to $\approx 20 \, a$ 
(solid and dashed lines as in Figs. \ref{proj.hern.iso} and \ref{proj.hern.aniso}). 
The weights were derived from $f_\mathrm{pl}$ via equation~(\ref{weights}). The kinematics 
along the major and the minor axis have been calculated from higher order Jeans equations 
\citep{mag94}. Before determining the GH-parameters the projected moments were integrated
along a $3.6$ arcsec wide major-axis slit and a $2.0$ arcsec wide minor-axis slit.
(Note that for the axisymmetric case we take $\beta = 1 - \sigma_\vartheta^2/\sigma_r^2$.)

As in the spherical case the Gauss-Hermite parameters of the projected kinematics
are reproduced to better than a few percent. Deviations in the outer parts of the
$H_4$-profile stem from the GH-fit and are not seen in the LOSVDs. 
The density distribution is also well reproduced up to $\approx a/10$ and the
anisotropy parameter is $|\beta| < 0.1$ from the outer edge of the library down to
$\approx a/10$.

\subsection{Changing the spatial coverage of the library}
\label{spatialproj}
The library only discretely represents a finite part of the available phase-space.
To check how this affects the accuracy of the calculation of phasespace integrals
of a given DF with the library, we did the
profile comparisons described in Sects. \ref{osmproj} and \ref{plumproj} for
libraries with different spatial extent and for different resolutions
in the space of orbital integrals.

\begin{figure}
\includegraphics[width=84mm,angle=0]{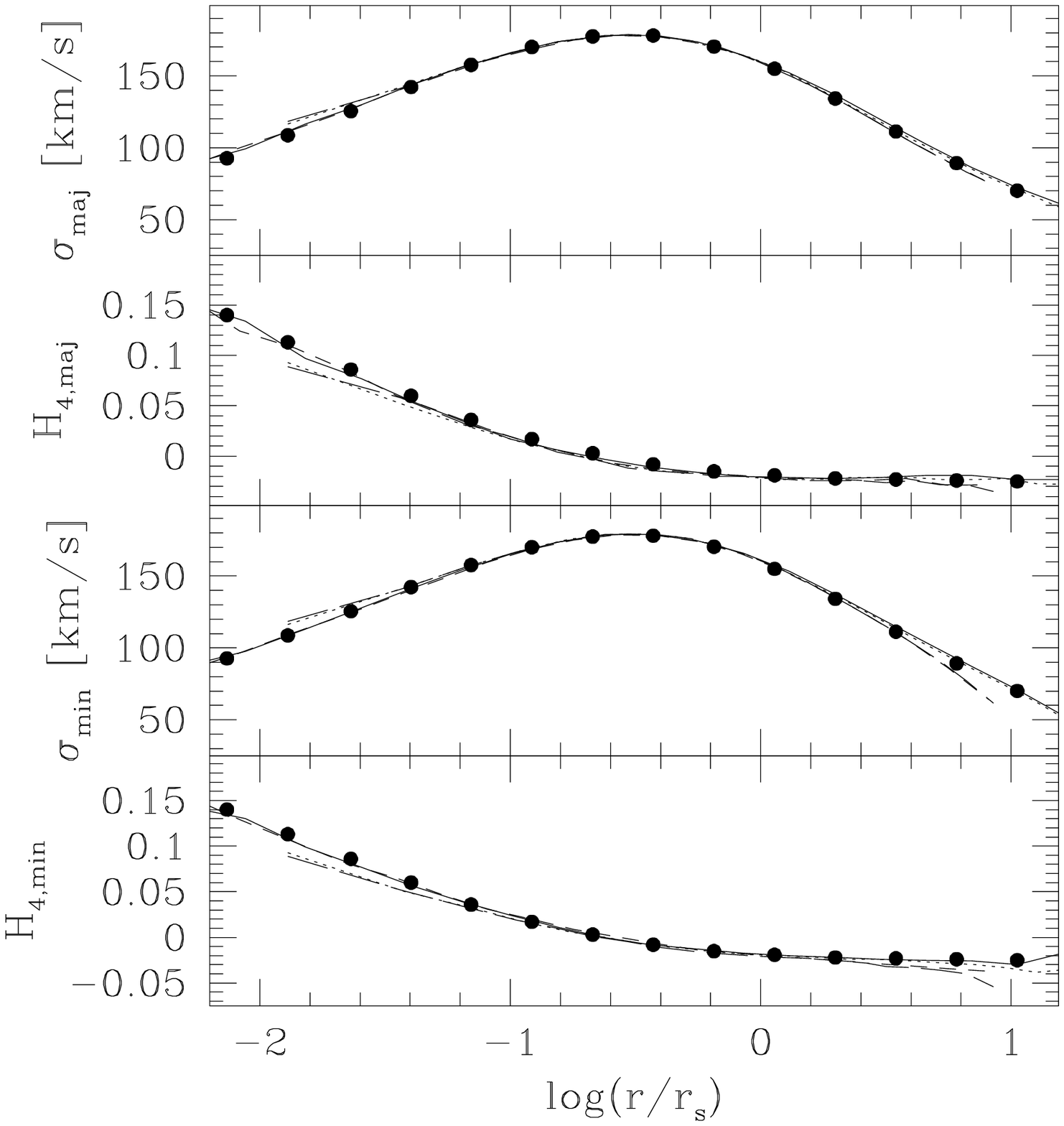}
\includegraphics[width=84mm,angle=0]{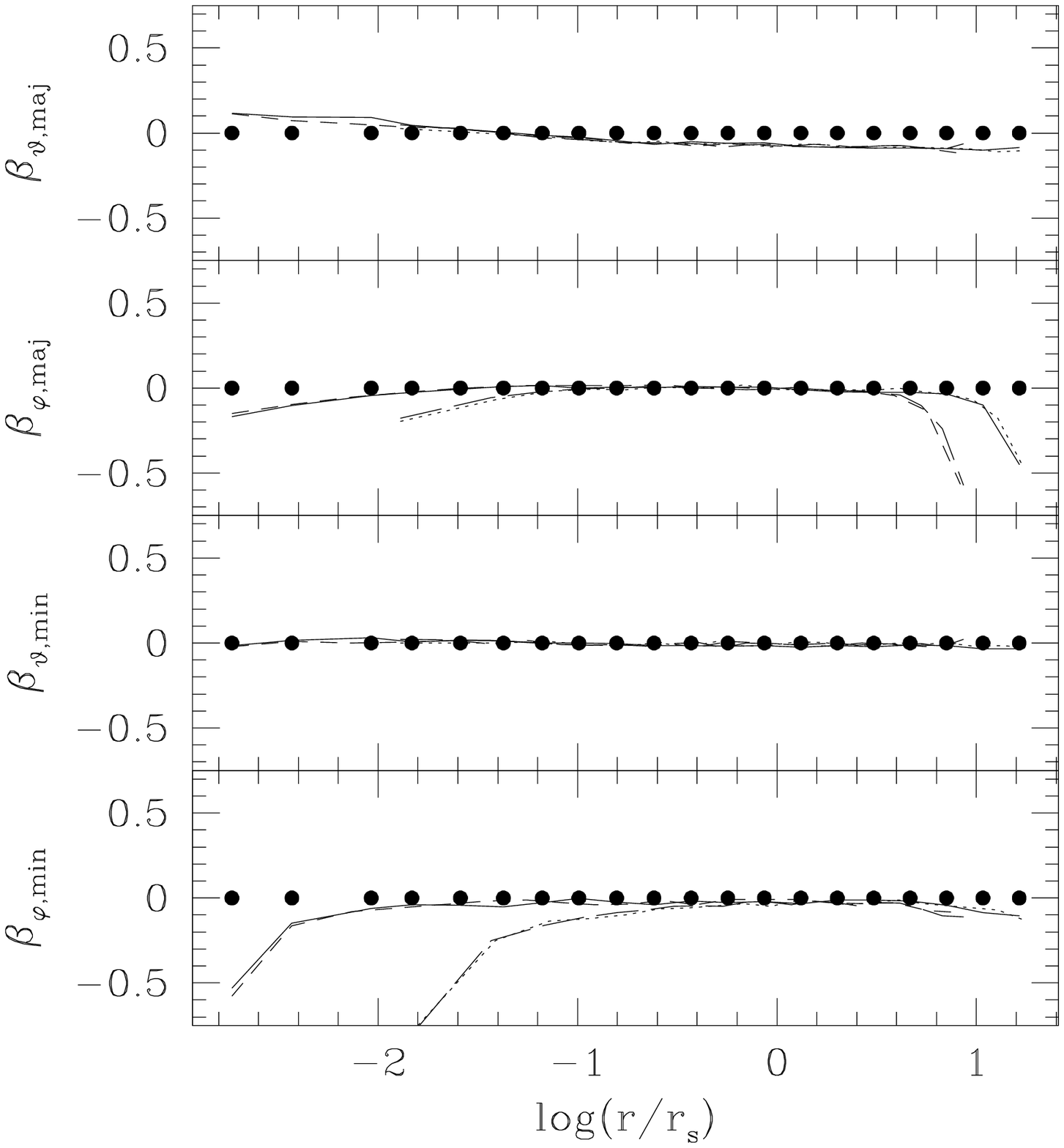}
\caption{$\sigma$ and $H_4$ (upper panel) and anisotropy (lower panel) 
along major and minor axis for the
isotropic Hernquist model (big dots) and four libraries with different
spatial extension ($r_\mathrm{min}/r_\mathrm{eff},r_\mathrm{max}/r_\mathrm{eff}$): 
($2.5 \times 10^{-4},10$) solid line, ($2.5 \times 10^{-3},10$) dotted line,
($2.5 \times 10^{-4},5$) short dashed line, ($2.5 \times 10^{-3},5$) long dashed line.}
\label{rminrmax}
\end{figure}

The upper panel of Fig.~\ref{rminrmax} shows $\sigma$ and $H_4$ along the
major and the minor axis for the isotropic Hernquist model (big dots). 
The four different lines show the outcome of four
libraries with different spatial coverage. For the solid line  
($r_\mathrm{min}=2.5 \times 10^{-4}, r_\mathrm{max}=10$) (in units of the effective radius), 
for the dotted line ($r_\mathrm{min}=2.5 \times 10^{-3}, r_\mathrm{max}=10$), for
the short dashed line ($r_\mathrm{min}=2.5 \times 10^{-4}, r_\mathrm{max}=5$) 
and for the long dashed line ($r_\mathrm{min}=2.5 \times 10^{-3}, r_\mathrm{max}=5$).

As expected the less extended libraries fail to reproduce the innnermost or outermost 
datapoins, respectively. In the vicinity of the equatorial plane (e.g. along the major
axis and at the central parts of the minor axis) the library becomes dominated by
azimuthal motion, when approaching $r_\mathrm{min}$ or $r_\mathrm{max}$ ,
 since orbits coming from further outside or inside are missing. Consequently,
the LOSVDs are too flat (too small $H_4$) as compared to the expectations (see e.g. the
outermost parts of the dashed lines of the libraries with small $r_\mathrm{max}$ along
the major axis and the innermost parts of the long-dashed and pointed lines of the libraries 
with large $r_\mathrm{min}$ in the minor-axis $H_4$-profile).

The effect can also be seen in the internal dynamical structure, which is illustrated in the 
lower panel of Fig.~\ref{rminrmax} where the anisotropy of the library with respect to
$\varphi$ and $\vartheta$ is plotted separately,
\begin{equation}
\beta_\varphi \equiv 1 - \frac{\sigma_\varphi^2}{\sigma_r^2}, \;
\beta_\vartheta \equiv 1 - \frac{\sigma_\vartheta^2}{\sigma_r^2}.
\end{equation}
Near the centre $\beta_\varphi < 0$ along the major and minor axis, confirming
the dominance of $\varphi$-motion brought about by the dominance of orbits
having their inner turning points there and consequently
rotate fastly around the axis of symmetry. The effect is less pronounced at the outer points
of the major axis, where the effective potential of the meridional plane motion is less
dominated by the $L_z$-term.

The $\beta_\vartheta$-profiles lack from boundary effects because they are independent from 
the ($E,L_z$)-sampling and simply reflect the degree to which the SOSs are filled with
orbital invariant curves.

Along the minor axis the agreement of library and model in projected $\sigma$
is quite good. Near the centre the library's $\sigma$ is enhanced due to the orbits
having their pericenter there.

\subsection{Changing the number of orbits in the library}

\begin{figure}
\includegraphics[width=84mm,angle=0]{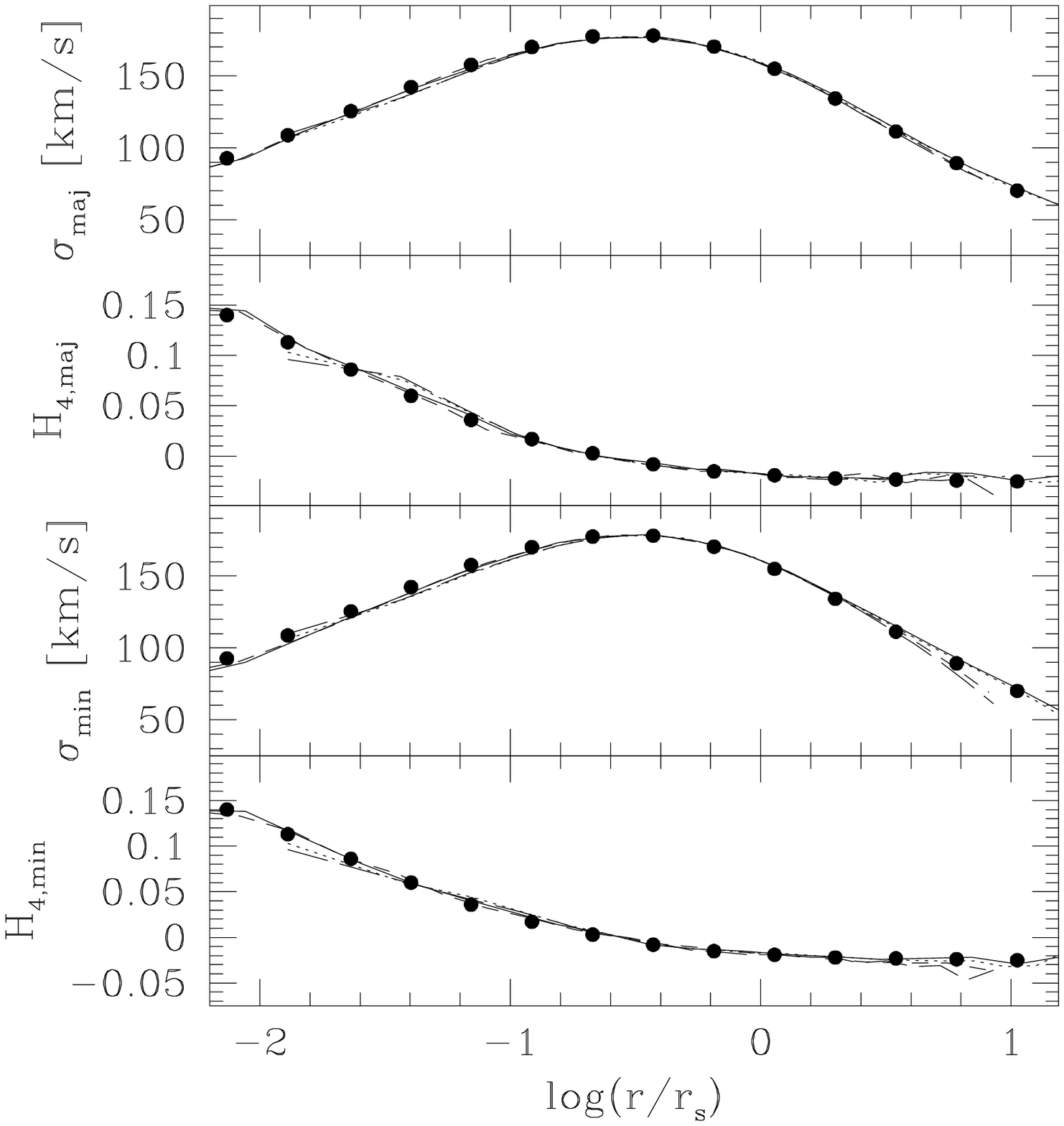}
\includegraphics[width=84mm,angle=0]{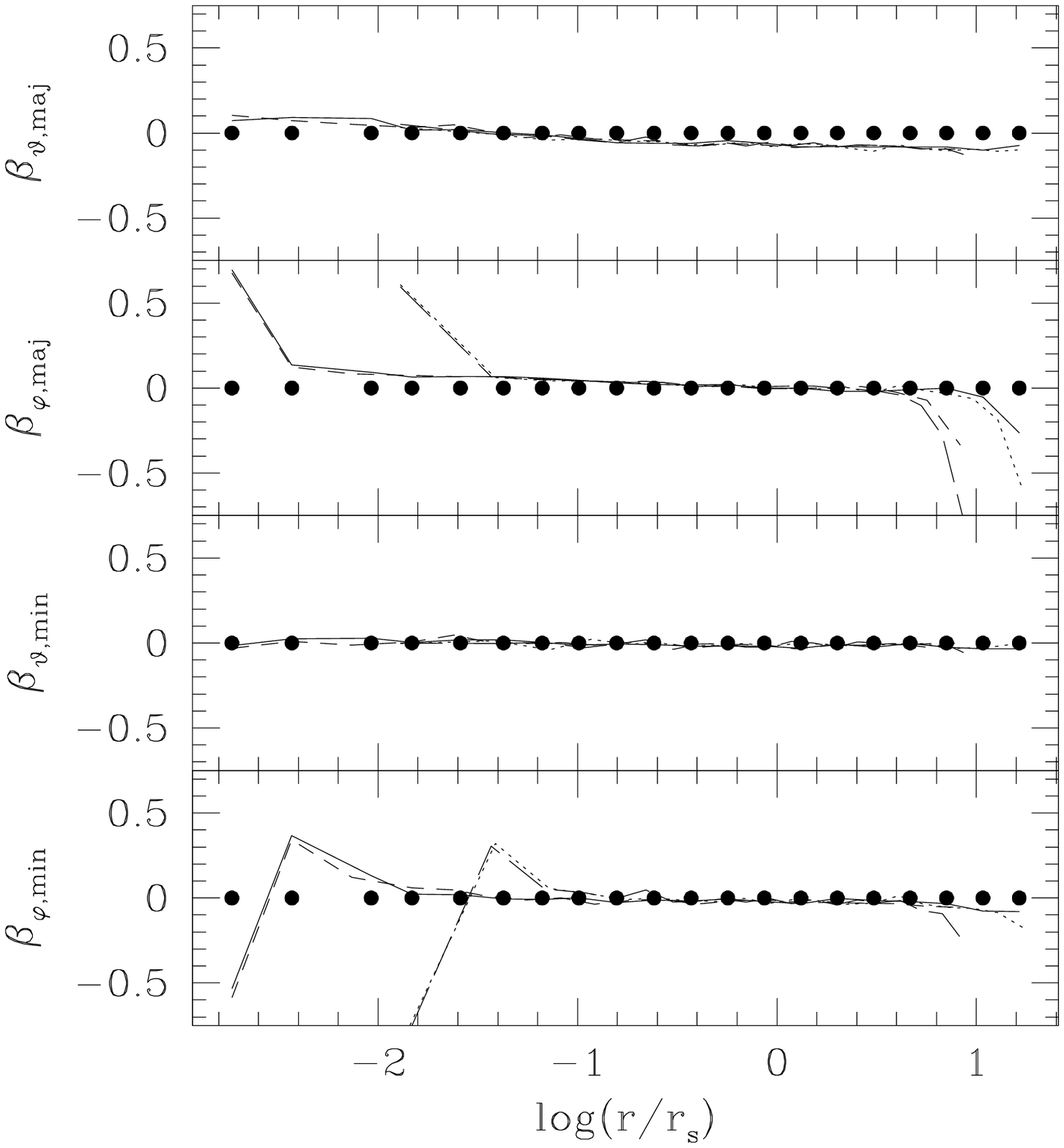}
\caption{Same profiles as in Fig.~\ref{rminrmax}, but the libraries have been set up with
a coarser sampling with roughly half the number of orbits as compared to 
Fig.~\ref{rminrmax}.}
\label{rminrmax2}
\end{figure}

Fig.~\ref{rminrmax2} shows the same comparison as Fig.~\ref{rminrmax}, but
for libraries, where we have skipped every second $r_\mathrm{peri}$ resulting
in only $\approx 2 \times 4700$ orbits per library. The gross appearance of 
Fig.~\ref{rminrmax2} is quite similar to Fig.~\ref{rminrmax} with some
minor differences. First, the scatter in the GHPs has increased a bit, however
the match of predictions and library is still on a level of a few percent.

The most striking difference is the increase of radial relative to azimuthal motion 
near the centre of the library.
Most probably this reflects the fact that the pericenters of the orbit sequences are located
at the inner edge of each radial bin. Therefore the most radial orbits which contribute
also significantly to $v_\phi$ near their turning points move through the whole bin
before turning around and thus rise the radial velocity dispersion. This effect is
strongest in the centre since our binning there becomes relatively large compared to 
the variation of the potential.
The balance between radial and meridional motion is not affected by this
resolution-effect, because the sampling {\it inside} each sequence (in the SOS) is 
independent from the ($E,L_z$)-grid and thus independent from the resolution of the
sampled peri- and apocenters.

\section[]{Fitting the library}
\label{libfit}
So far we have omitted the problem of finding the orbital weights
$w_i$ according to some given kinematical constraints. This section
contains a brief description of our use of the maximum entropy technique of
\citet{maxs} to fit the library to some LOSVDs.

\subsection{Maximum entropy technique}
\label{maxent}
Given a set of kinematic constraints, we seek the
orbital weights, that best fit the library to the constraints. These weights
are derived from the maximization of an entropy-like quantity \citep{maxs}
\begin{equation}
\label{entropy}
\hat{S} \equiv S - \alpha \chi^2,
\end{equation}
where 
\begin{equation}
\label{chilosvd}
\chi^2=\sum_{j,k} 
\left(
\frac{\mathrm{LOSVD(lib)}^{jk}-\mathrm{LOSVD(data)}^{jk}}
{\Delta \mathrm{LOSVD(data)}^{jk}}
\right)^2
\end{equation}
gives the departure between the predicted kinematics
of the library $\mathrm{LOSVD(lib)}$ (cf. equation~(\ref{liblosvd})) 
and the data kinematics $\mathrm{LOSVD(data)}$ . Note that the luminosity density 
$\nu$ is not fitted, but used as a boundary condition (see Richstone et al. (in preparation) 
for details). $S$ is an approximation to the usual Boltzmann-entropy
\begin{equation}
\label{bentropy}
S \equiv \int f_\mathrm{lib} \ln \left( f_\mathrm{lib} \right) \,
\der^3r \, \der^3v
\, = \sum_i w_i \ln \left( \frac{w_i}{V_i} \right).
\end{equation}

In the absence of any other condition the maximization of $S$ enforces the
weights $w_i$ to be proportional to the phase-volumes $V_i$. This fact can be used to bias
the library towards any set of predefined weights. If, for example, we substitute the 
phase-volumes in equation~(\ref{bentropy}) by $V_i \rightarrow f_i \, V_i$,
\begin{equation}
\label{bentropybias}
S \rightarrow S' = \sum_i w_i \ln \left( \frac{w_i}{f_i \, V_i} \right),
\end{equation}
then the maximization of $S'$ yields weights $w_i$ proportional to $f_i \, V_i$. According
to equation~(\ref{weights}) one can choose the factors $f_i$ to bias the library towards
any given DF $f$. The Boltzmann entropy corresponds to the case of equal a priori 
probabilities $f_i = f_j$ in phase-space.

\subsection{The smoothing parameter $\alpha$}
The smoothing parameter $\alpha$ controls the influence of the entropy $S$ on the
fitted weights. If $\alpha$ is small, then the maximum of $\hat{S}$ is
less affected by $\chi^2$ and the library
gives a poor fit to the data. Consequently, it will not represent the true structure
of the object to which it is fitted. If on the other hand $\alpha$ is large, then
the maximum of $\hat{S}$ is mostly determined by the minimum of $\chi^2$. In this
case the library fits the noise in the data. The DF of the library is then highly unsmooth
and again does not represent the true DF of the corresponding object.

The question of how much smoothing has to be applied in order to get an optimal
estimate of the true underlying DF for a given set of observational data with specific
errors and spatial sampling will be the content of a forthcoming paper. Here,
we focus on illustrating the accuracy of our method to setup the orbit
libraries. In the following we will always show the results for that $\alpha$ which
gives the best match to the input DF.


\section[]{Reconstructing distribution functions from fitted libraries}
\label{dfrecon}
In this Section we use the DFs of Sec.~\ref{projdf}, but instead of exploiting 
equation~(\ref{weights}) to {\it assign} the orbital weights and to compare spatial
profiles of the library and the original DF, we now fit the library to the DF as
follows. First, we calculate the density profile and GHPs connected with the DF
\begin{equation}
\rho = \int f \, \der^3 v
\end{equation}
and
\begin{equation}
\mathrm{LOSVD}_f(v_{los}) = \frac{1}{\rho} \int f \, \der^2 v_{\bot}
\end{equation}
where the GHPs are obtained from the LOSVDs as described in Sec.~\ref{osmproj}. We
compose a library as dscribed in Sec.~\ref{library} and fit it to the GHPs via
the maximum entropy technique of Sec.~\ref{libfit}. Finally, we compare the orbital weights
$w_i(\alpha)$ resulting from the fit with those expected from the DF via equation~(\ref{weights}).
By showing that the fitted weights approximate the input DF over a large region in
phase-space, we justify that we can use the degree to 
which the library approximates the DF as a criterium to determine the optimal amount of
smoothing, which we will exploit in a subsequent paper in more detail.

In order to find the best fit weights that minimize the $\chi^2$ of Eq.(\ref{chilosvd}), we
derived error bars for the LOSVDs by first assigning error bars 
to the GHPs and then determining LOSVD errors by means of Monte-Carlo simulations.
The error for $\sigma$ was choosen to linearily increase with $r$ from 2 per cent at the 
innermost data point to 10 per cent for the outermost data point. For $H_3$ and $H_4$ the
errors increase from $0.01$ to $0.05$. The definition of the errors is somewhat arbitrary
since we do not add noise to the data points, but they are roughly comparable to
real data error bars. Since $v = 0$ in the models, the error for $v$ is set to 
$\Delta v = 2 \ \mathrm{km} \, \mathrm{s}^{-1}$. 
A detailed investigation of the influence of realistic errors on 
the accuracy of the reconstructed internal properties of a fitted library will be presented 
in a forthcoming paper.

\subsection{Hernquist model}

\begin{figure}
\includegraphics[width=84mm,angle=0]{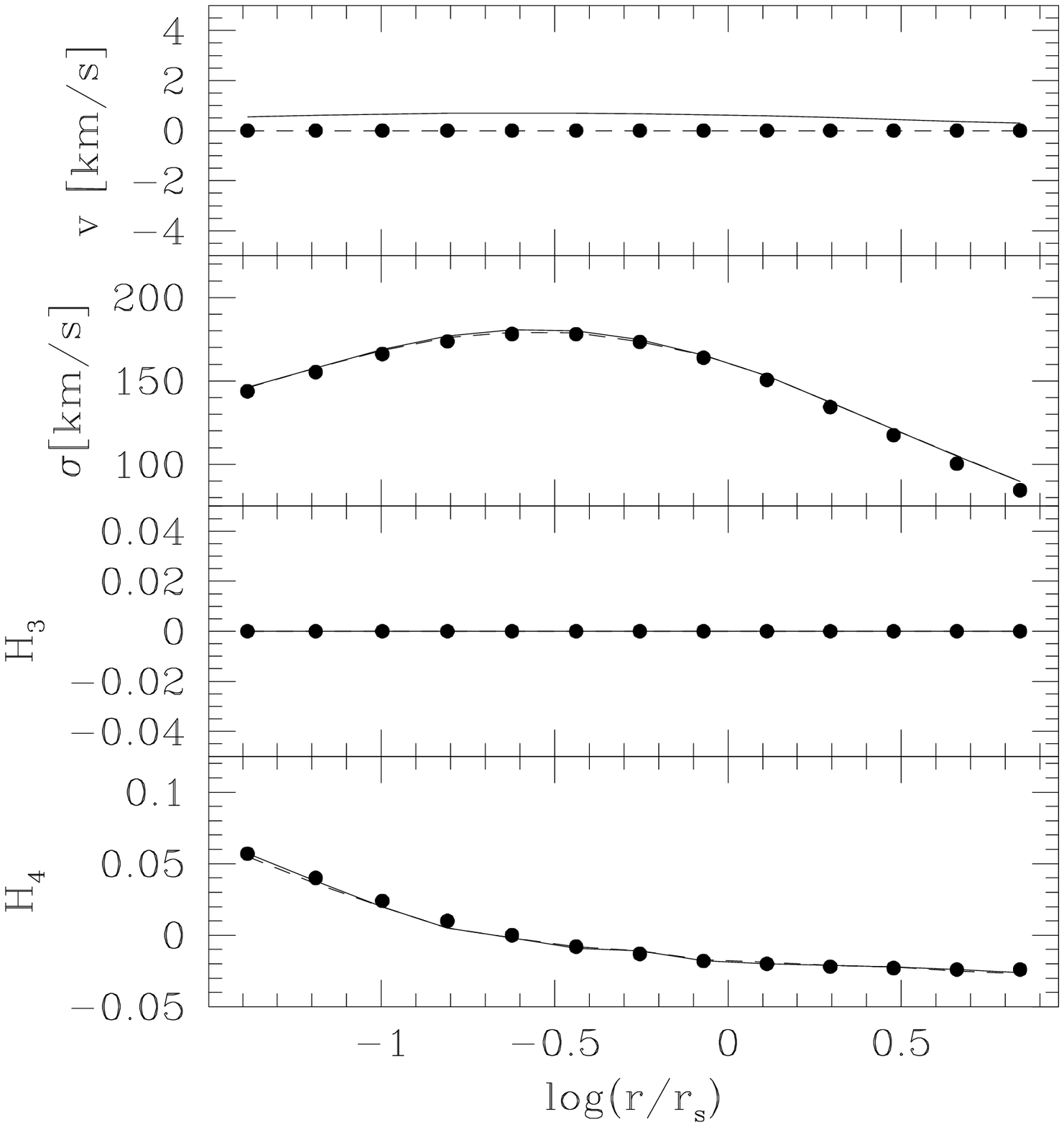}
\includegraphics[width=84mm,angle=0]{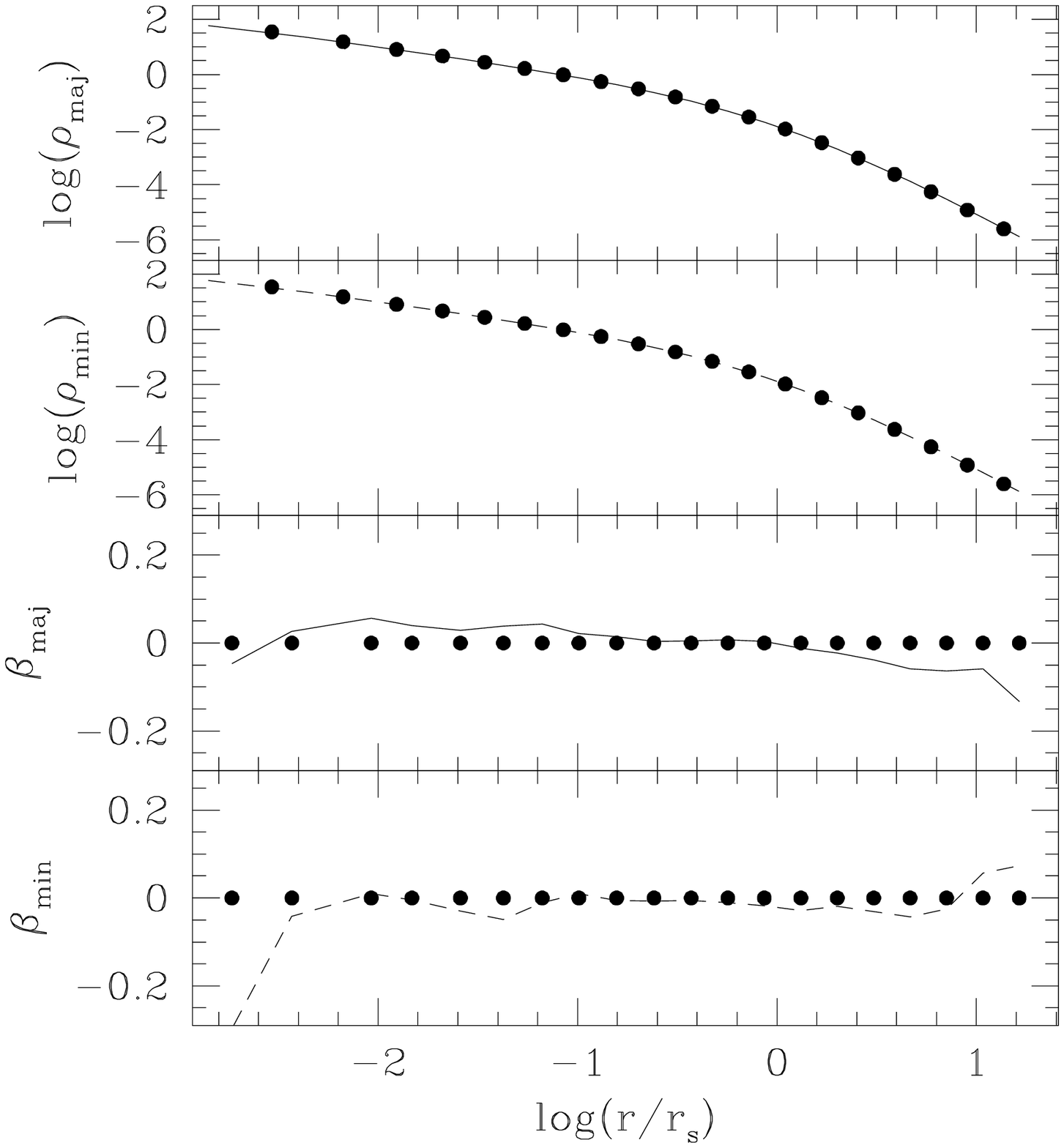}
\caption{Comparison of a library fitted to the LOSVDs of a spherical, isotropic
Hernquist DF (lines) and the Hernquist model itself (big dots).
The upper panel shows 
the projected kinematics along the major axis (solid line) and minor axis (dashed
line). The lower panel shows the density distribution (upper two rows,
$[\rho] = \mathrm{M}_{\sun} \, \mathrm{pc}^{-3}$) and the
anisotropy parameter (lower two rows) for the two axes.}
\label{fit.hern.iso}
\end{figure}

Fig.~\ref{fit.hern.iso} shows a comparison of characteristic properties of a library
fitted to the kinematics corresponding to the dots in the upper panel and the original 
DF. The definition of the lines and dots as well as the input DF
are the same as for Fig.~\ref{proj.hern.iso}, the fit was obtained with
$\alpha=0.0046$. As expected the match to the kinematics and the internal
density profile is excellent after the fit. The anisotropy is smaller than
$\left | \beta \right | < 0.1$ over a spatial region largely exceeding the area where the
LOSVDs were fitted. Only near the very centre, the minor-axis $\beta$-profile drops
significantly because of the lack of radial orbits coming from inside the inner
boundary of the library (cf. Sec.~\ref{projdf}).

\begin{figure}
\includegraphics[width=60mm,angle=-90]{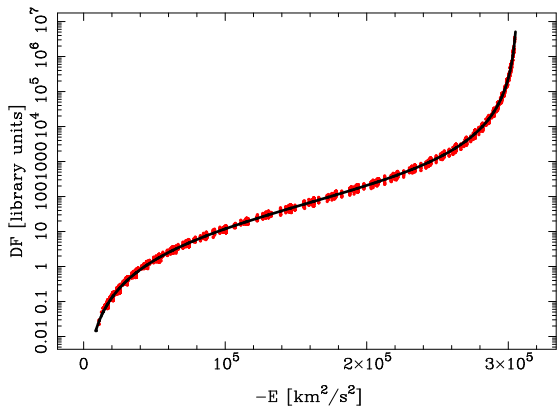}
\caption{Comparison of the DF of a spherical Hernquist model 
(solid line, units defined in the text) with the phase-space densities obtained from a 
library fitted to GHPs along two perpendicular axes in the galaxy (details in the text). 
Each dot represents a single orbit. The rms between library and model is $12.1$ per
cent over a region covering 90 per cent of the library's mass.}
\label{dfcomphern}
\end{figure}

Fig.~\ref{dfcomphern} shows the DF reconstructed from the fitted weights 
via equation~(\ref{orbdf}) (dots) together with the input DF (thick line).
Each dot represents the phase-space density along one single orbit, the
densities are scaled according to $\sum w_i = \sum V_i = 1$ . Over a region 
covering 90 per cent of the library's mass, the rms-difference between the Hernquist DF 
and the orbital phase-space densities is $12.1$ per cent. 
The remaining departures between model
and fit are mostly due to boundary effects arising from the discrete and finite nature
of the library.

\begin{figure}
\includegraphics[width=60mm,angle=-90]{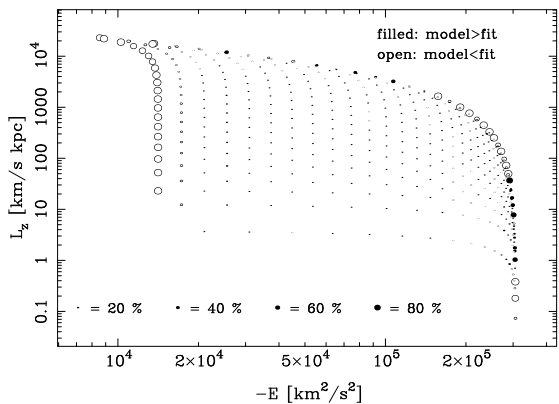}
\caption{The fractional difference between a spherical Hernquist model and
a fitted library as a function of orbital energy $E$ and z-angular
momentum $L_z$. Larger dots corresponds to larger differences. For open dots, the DF 
of the library overestimates the real DF, for the solid dots it underestimates the DF.}
\label{dfcomphernelz}
\end{figure}

Fig.~\ref{dfcomphernelz} shows the fractional differences of input model and library as
a function of orbital energy $E$ and angular momentum $L_z$. For each dot,
the contributions of individual orbits with common $E$ and $L_z$ have been integrated.
Larger dots correspond to larger differences between input DF and fitted library.
From Fig.~\ref{dfcomphernelz} one sees that the remaining deviations between library
and input DF mostly stem from orbits lying at the boundary of the phase-space-region
covered by the library. Since the library only contains a finite number of all orbits,
the fit to the kinematics with the density as a boundary condition enforces some
redistribution of orbits as compared to the original  DF. For example,
at the outer boundary of the library ($E \approx 0$) the fitted orbital phase-space
densities are too large as compared to the input DF. These orbits compensate
the cut-off in energy and contain all the light that should have been
distributed along even lower bound orbits. For the same reason, the library
fails to reproduce the Hernquist DF near the most bound orbits.

\begin{figure}
\includegraphics[width=84mm,angle=0]{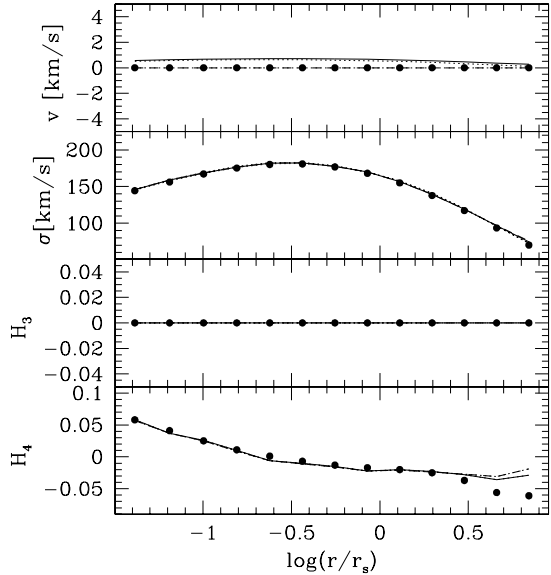}
\includegraphics[width=84mm,angle=0]{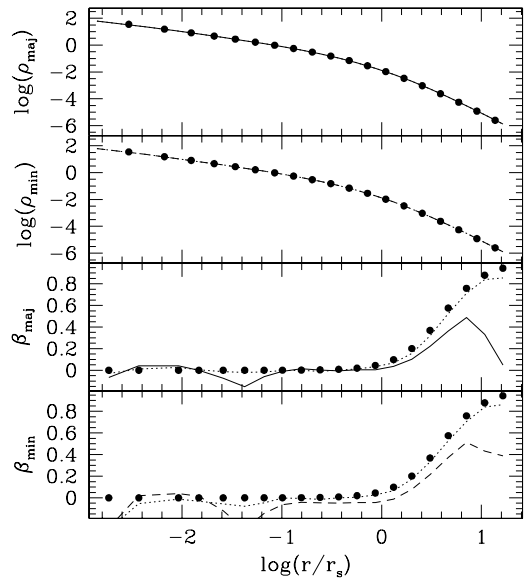}
\caption{Same as Fig.~\ref{fit.hern.iso} but for an anisotropic Hernquist model
with $r_a = 4 \, r_s$. The dotted line shows the result of a fit with ``biased weights'' 
(see text for details).}
\label{fit.hern.aniso}
\end{figure}

Fig.~\ref{fit.hern.aniso} shows the results when fitting the same library to the projected
kinematics of the anisotropic Hernquist model with $r_a = 4 r_s$, corresponding to the dots
in the upper panel of the Figure. Again, after the fit the library perfectly
reproduces the internal density profile and the projected kinematics.
The mismatch in the outer parts of the $H_4$-profiles result from errors in the GHP-fit 
(cf. Sec.~\ref{osmproj}). However, we don't fit the library to the GHP, but directly to the 
LOSVD. The $\beta$-profiles of the library
follow the expected curves well inside the region covered with kinematical constraints.
In the outer parts however they do not follow the input model to predominant radial
motion but turn back to an isotropic appearance. This is a reflection of the entropy
maximization used in the fit, which forces those parts of the library to isotropy, which
are not constrainted by data points. 

To confirm that, we refitted the library, however
replaced the $V_i$ in equation~(\ref{bentropy}) by the weights of the anisotropic Hernquist DF
following from equation~(\ref{weights}). Since for the maximum entropy solution of 
equation~(\ref{bentropy}) (without any other condition) the weights $w_i$ are proportional to the
values $V_i$, now being the weights of the anisotropic DF instead of the phase-volumes, this 
biases the fit towards the anisotropic Hernquist model. The characteristics of the 
corresponding fit are displayed by the dotted lines in Fig.~\ref{fit.hern.aniso}. 
The projected kinematics and internal density are indistinguishable from the maximum entropy 
fit, but now the anisotropy profile is in perfect agreement with the input model.

\subsection{Flattened Plummer model}

\begin{figure}
\includegraphics[width=84mm,angle=0]{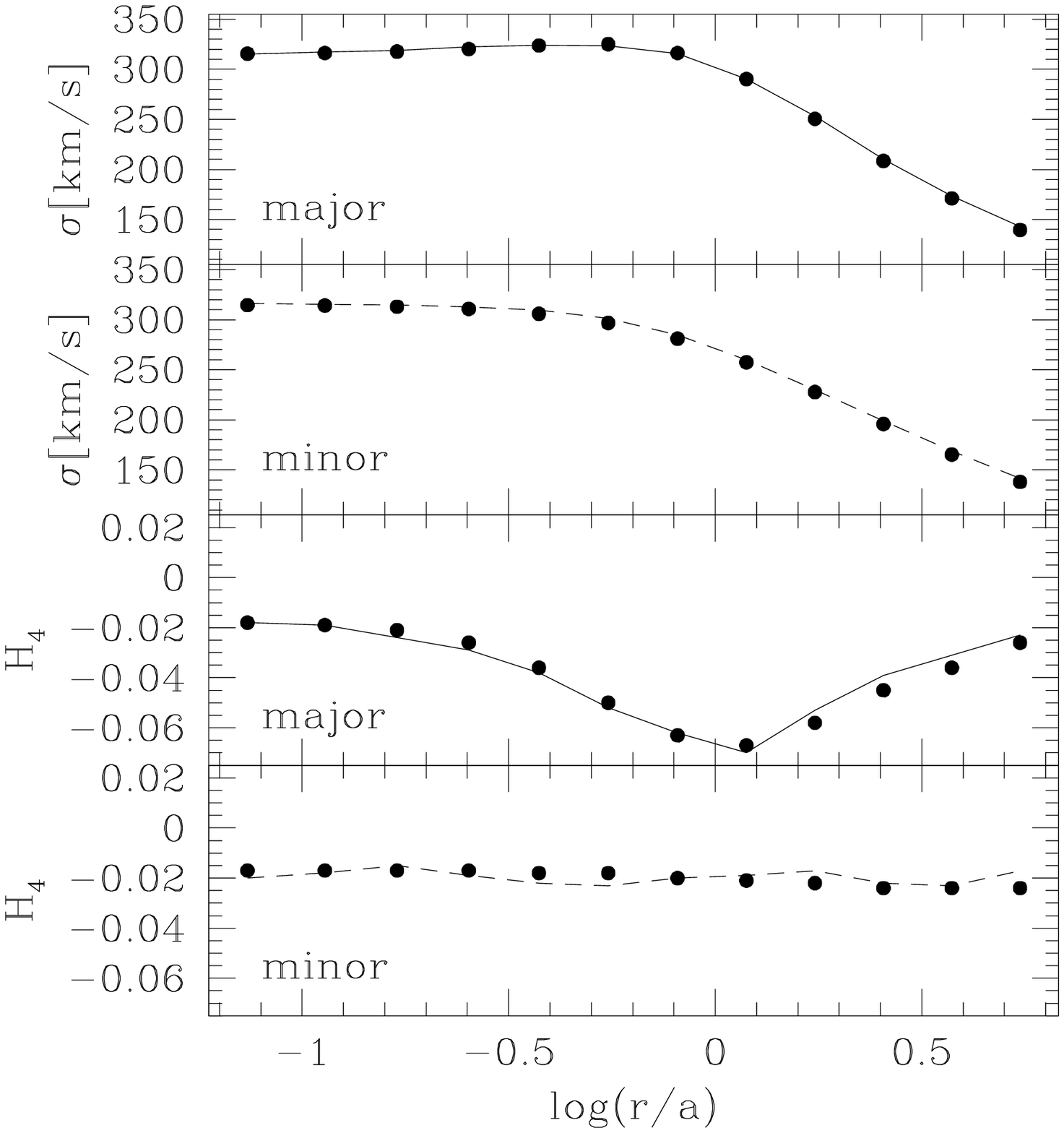}
\includegraphics[width=84mm,angle=0]{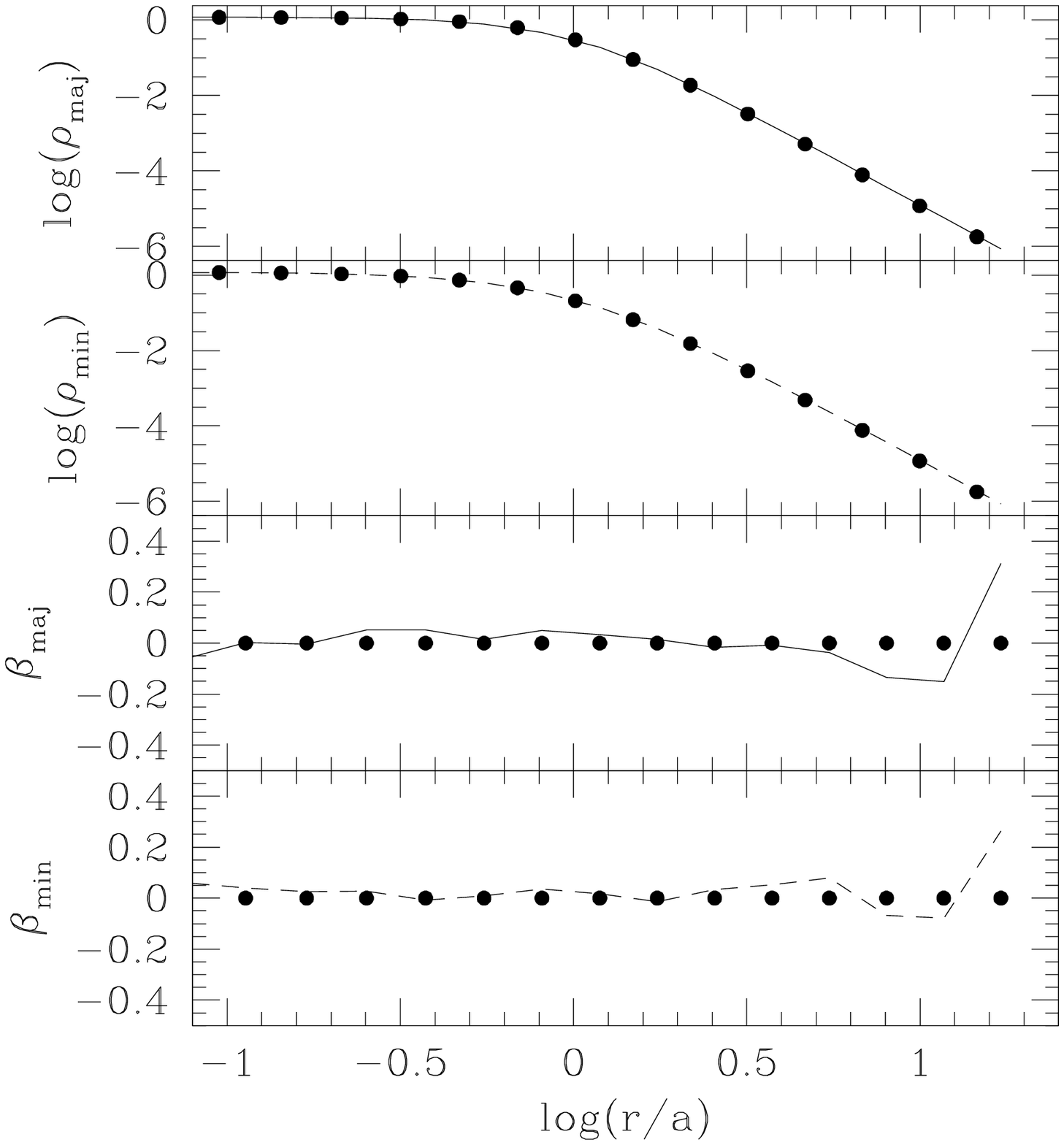}
\caption{Comparison of a flattened Plummer model (big dots) and a fitted library 
(lines). The upper panel shows the projected kinematics along the major axis (solid lines) 
and minor axis (dashed lines). The lower panel shows internal moments along the minor 
and major axis, respectively (units as in Fig.~\ref{proj.plummer}).}
\label{fit.plummer}
\end{figure}

Fig.~\ref{fit.plummer} shows the GHPs and internal density and anisotropy of the
Plummer model with $b = a/2$ of Sec.~\ref{plumproj} together with a fitted library containing 
$\approx 2 \times 4400$ orbits. The library was fitted to the LOSVDs corresponding to
the dots of the upper panel of the Figure with a 
smoothing parameter of $\alpha \approx 0.03$. The small
deviations between the library's kinematics and the model in the upper panel of the
figure are due to the low resolution in the GH-fit and are not seen in the LOSVDs 
used for the fit. The anisotropy parameter is confined to $|\beta| < 0.1$ all over
the region where the library is constrained by kinematic data.

\begin{figure}
\includegraphics[width=60mm,angle=-90]{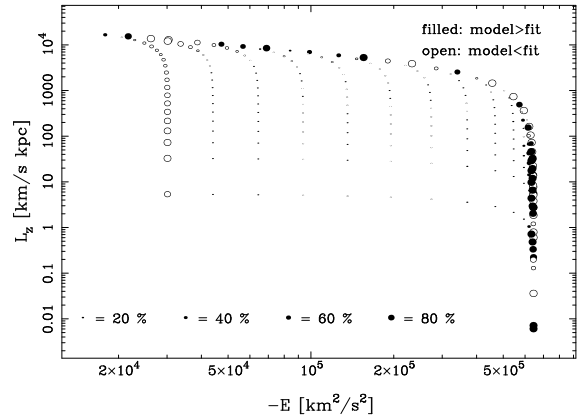}
\caption{Deviations from the reconstructed DF of a fitted library and the Plummer DF of
Fig.~\ref{fit.plummer}. Each dot represents one sequence of orbits with common $E$ and 
$L_z$. For the open dots, the DF of the library overestimates the real DF, for the
solid dots it underestimates it. Larger dots indicate larger differences.}
\label{dfcompplumelz}
\end{figure}

The rms-difference between the reconstructed DF and the input model is $\approx 15$ per
cent over a region covering 90 per cent of the library's mass. 
As Fig.~\ref{dfcompplumelz} shows,
differences between model and library are confined to the boundaries of the sampled
($E,L_z$)-region of the phase-space. As for the Hernquist model, the reason for these 
differences is the incomplete orbit sampling at the edges of the library.

\section[]{Summary}
\label{conclusion}
We have presented a modified version of the Schwarzschild code of Richstone et al. (in
preparation). The code involves a new orbit sampling at given energy $E$ and angular momentum
$L_z$ and a new implementation for the calculation of the orbital phase-volumes.

For our libraries we supplement the drop of orbits with common energy $E$ and
angular momentum $L_z$ from the ZVC as described in Richstone et al. (in
preparation) by scanning the SOS
with a resolution that varies as the sampling progresses from the more radial to the
more shell-type orbits. This sampling has been shown to completely
fill the SOS connected with a pair ($E,L_z$) with orbital imprints.

A Voronoi tessellation of the SOSs of orbits with common $E$ and $L_z$ allows us to
calculate the phase-space-volumes of individual orbits in any axisymmetric potential. With
the phase-volumes we can convert the orbital weights describing the relative contribution
of the orbits to the whole library into phase-space densities and vice versa. As a first 
application we use the densities to check our method to setup the library in two different
ways.

First, we calculate spatial profiles of internal and projected properties of isotropic 
and anisotropic DFs of spherical $\gamma$-models as well as of the flattened Plummer model 
with the library. The density profiles,
anisotropy profiles and projected kinematics of the library closely match those directly
inferred from the corresponding DF. The errors in the higher order GH-parameters $H_n$
are $\Delta H_n < 0.01$, for $n=3,4$ and the fractional error in the projected dispersion
$\Delta \sigma < 1$ per cent are accurate on a level better than that of present 
day observational
errors. The largest deviations occur in the anisotropy profile, but are smaller than
$\Delta \beta < 0.1$ at almost all positions in the library, however increase towards 
the edges of the spatial region which is covered by the orbits. This boundary effect is
caused by the locally incomplete orbit sampling there. If in
practical applications the libraries are constructed to extend beyond the area with 
observational constraints, then these inaccuracies are negligible. Hence, our libraries 
fairly represent the phase-space structure of the considered models.

As a second application we {\it fitted} libraries to the GHPs of the same spherical 
$\gamma$-models and flattened Plummer models. The reconstructed DFs match the
input DF with a rms of about 15 per cent over a region covering 90 per cent of the 
library's mass. The remaining deviations are mostly restricted to orbits at the boundary 
of the phase-space volume represented by the library. This is not unexpected since the library
only discretely represents a finite subregion of the input system. Consequently some
redistribution of orbits is necessary to compensate for orbits not included in the library.

We will investigate the influence of observational errors on the reconstructed
DFs and of the amount of smoothing applied in the fit in a forthcoming publication.
In a further step we will reconstruct the internal structure and mass composition of a sample
of flattened early-type galaxies in the axisymmetric approximation.

\section*{Acknowledgments}
J. Thomas acknowledges financial support by the 
Sonderforschungsbereich 375 ``Astro-Teilchenphysik'' of the Deutsche Forschungsgemeinschaft.

\bsp

\label{lastpage}


\begin{thebibliography}{99}
\bibitem[\protect\citeauthoryear{Binney \& Mamon}{1982}]{BM82} Binney J., Mamon G.~A., 1982, MNRAS, 200, 361
\bibitem[\protect\citeauthoryear{Binney, Gerhard \& Hut}{1985}]{binney85} Binney J., Gerhard O.~E., Hut P., 1985, MNRAS, 215, 59 
\bibitem[\protect\citeauthoryear{Binney \& Tremaine}{1987}]{binneytremaine} Binney J., Tremaine S., 1987, Galactic Dynamics (princeton: Princeton University Press)
\bibitem[\protect\citeauthoryear{Cappellari et al.}{2002}]{cappellari02} Cappellari M., Verolme E.~K., van der Marel R.~P., Verdoes Kleijn G.~A., Illingworth G.~D., Franx M., Carollo C.~M., de Zeeuw P.~T., 2002, ApJ, 578, 787
\bibitem[\protect\citeauthoryear{Carollo, de Zeeuw \& van der Marel}{1995}]{carollo95} Carollo C.~M., de Zeeuw P.~T., van der Marel R.~P., 1995, MNRAS, 276, 1131 
\bibitem[\protect\citeauthoryear{Cretton et al.}{1999}]{cretton99} Cretton N., de Zeeuw P.~T., van der Marel R.~P., Rix H.~W., 1999, ApJS, 124, 383 
\bibitem[\protect\citeauthoryear{Cretton \& Emsellem}{2004}]{cretton04} Cretton N., Emsellem E., 2004, MNRAS 347, L31
\bibitem[\protect\citeauthoryear{Dehnen \& Gerhard}{1993}]{DG93} Dehnen W., Gerhard O.~E., 1993, MNRAS, 261, 311 
\bibitem[\protect\citeauthoryear{Dehnen}{1993}]{D93} Dehnen W., 1993, MNRAS, 265, 250 
\bibitem[\protect\citeauthoryear{Dejonghe \& Merritt}{1992}]{DM92} Dejonghe H., Merritt D.~R., 1992, MNRAS, 391, 531 
\bibitem[\protect\citeauthoryear{Gebhardt et al.}{2003}]{G03} Gebhardt K.~et al., 2003, ApJ, 583, 92 
\bibitem[\protect\citeauthoryear{Gerhard}{1993}]{Ger93} Gerhard O.~E., 1993, MNRAS, 265, 213
\bibitem[\protect\citeauthoryear{Hernquist}{1990}]{hernquist} Hernquist L., 1990, ApJ, 356, 359 
\bibitem[\protect\citeauthoryear{Lynden-Bell}{1962}]{LB62} Lynden-Bell D., 1962, MNRAS, 123, 447 
\bibitem[\protect\citeauthoryear{Magorrian \& Binney}{1994}]{mag94} Magorrian J., Binney J., 1994, MNRAS, 271, 949
\bibitem[\protect\citeauthoryear{Mehlert et al.}{2000}]{coma1} Mehlert D., Saglia R.~P., Bender R., Wegner G., 2000, A\&AS, 141, 449
\bibitem[\protect\citeauthoryear{Merritt}{1985a}]{M85a} Merritt D.~R., 1985a, AJ, 90, 1023
\bibitem[\protect\citeauthoryear{Merritt}{1985b}]{M85b} Merritt D.~R., 1985b, MNRAS, 214, 25 
\bibitem[\protect\citeauthoryear{Merritt \& Saha}{1993}]{MS93} Merritt D.~R., Saha P., 1993, ApJ, 409, 75
\bibitem[\protect\citeauthoryear{Osipkov}{1979}]{O79} Osipkov L.~P., 1979, Pis'ma Astron. Zh., 55, 77
\bibitem[\protect\citeauthoryear{Richstone \& Tremaine}{1988}]{maxs} Richstone D.~O., Tremaine S., 1988, ApJ, 327, 82 
\bibitem[\protect\citeauthoryear{Richstone et al.}{2004}]{richstone04a} Richstone D.~O. et al, 2004, preprint, astro-ph/0403257
\bibitem[\protect\citeauthoryear{Rix et al.}{1997}]{R97} Rix H.~W., de Zeeuw P.~T., Cretton N., van der Marel R.~P., Carollo C.~M., 1997, ApJ, 488, 702 
\bibitem[\protect\citeauthoryear{Romanowsky \& Kochanek}{1997}]{roman97} Romanowsky A.~J., Kochanek C.~S., 2001, ApJ, 553, 722
\bibitem[\protect\citeauthoryear{Schwarzschild}{1979}]{S79} Schwarzschild M., 1979, ApJ, 232, 236 
\bibitem[\protect\citeauthoryear{Shewchuk}{1996}]{shew} Shewchuk J., 1996, in First Workshop on Applied Computational Geometry, ACM (http://www.cs.cmu.edu/\~{}quake/triangle.html) 124
\bibitem[\protect\citeauthoryear{Valluri, Merritt \& Emsellem}{2004}]{valluri04} Valluri M., Merritt D., Emsellem E., 2004, ApJ, 602, 66
\bibitem[\protect\citeauthoryear{van de Ven et al.}{2003}]{vandeven03} van de Ven G., Verolme E.~K., Cappellari M., de Zeeuw P.~T., 2003, in Dark Matter in Galaxies, IAU Symposium No. 220, eds. S. Ryder, D.J. Pisano, M. Walker and K. Freeman
\bibitem[\protect\citeauthoryear{van der Marel et al.}{1998}]{vdM98} van der Marel R.~P., Cretton N., de Zeeuw P.~T., Rix H.~W., 1998, ApJ, 493, 613
\bibitem[\protect\citeauthoryear{van der Marel \& Franx}{1993}]{vdMF93} van der Marel R.~P., Franx M., 1993, ApJ, 407, 525
\bibitem[\protect\citeauthoryear{Vandervoort}{1984}]{V84} Vandervoort P.~O., 1984, ApJ, 287, 475 
\bibitem[\protect\citeauthoryear{Verolme \& de Zeeuw}{2002}]{verolme02a} Verolme E.~K., de Zeeuw P.~T., 2002, MNRAS, 331, 959
\bibitem[\protect\citeauthoryear{Verolme et al.}{2002}]{verolme02b} Verolme E.~K. et al., 2002, MNRAS, 335, 517
\bibitem[\protect\citeauthoryear{Wegner et al.}{2002}]{coma2} Wegner G., Corsini E.~M., Saglia R.~P., Bender R., Merkl D., Thomas D., Thomas J., Mehlert D., 2002, A\&A, 395, 753
\end{thebibliography}
\end{document}